\documentclass[aps,prl,amsmath,twocolumn,amssymb,titlepage,superscriptaddress,floatfix,showpacs]{revtex4-1}
\usepackage{graphicx}
\usepackage{setspace}

\usepackage{xcolor}
\usepackage{ulem}

\usepackage{amsmath,amsthm,amssymb}
\usepackage{mathrsfs}

\usepackage{changes}

\newcommand{\ave}[1]{\langle #1\rangle}

\newcommand{\I}{\mathrm{i}}
\newcommand{\green}[2]{\langle\langle #1, #2 \rangle\rangle}

\newcommand{\beq}[1]{\begin{equation} #1 \end{equation}}
\newcommand{\bsplit}[1]{\begin{equation} \begin{split} #1 \end{split} \end{equation}}

\begin{document}
\title{Unveiling the underlying interactions in Ta$_2$NiSe$_5$ from photo-induced lifetime change}
\author{Denis Gole\v{z}}\thanks{These authors contributed equally.}
\affiliation{Jozef Stefan Institute, Jamova 39, SI-1000 Ljubljana, Slovenia}
\affiliation{Faculty of Mathematics and Physics, University of Ljubljana, Jadranska 19, 1000 Ljubljana, Slovenia}
\affiliation{Center for Computational Quantum Physics, Flatiron Institute, 162 Fifth Avenue, New York, NY 10010, USA}

\author{Sydney K. Y. Dufresne}\thanks{These authors contributed equally.}
\affiliation{Quantum Matter Institute, University of British Columbia, Vancouver, BC V6T 1Z4, Canada}
\affiliation{Department of Physics \& Astronomy, University of British Columbia, Vancouver, BC V6T 1Z1, Canada}

\author{Min-Jae Kim}
\affiliation{Max Planck Institute for Solid State Research, Heisenbergstrasse 1, D-70569 Stuttgart, Germany}

\author{Fabio Boschini}
\affiliation{Quantum Matter Institute, University of British Columbia, Vancouver, BC V6T 1Z4, Canada}
\affiliation{Department of Physics \& Astronomy, University of British Columbia, Vancouver, BC V6T 1Z1, Canada}
\affiliation{Centre \'{E}nergie Mat\'{e}riaux T\'{e}l\'{e}communications, Institut National
de la Recherche Scientifique, Varennes, Qu\'{e}bec J3X 1S2, Canada}

\author{Hao Chu} 
\affiliation{Quantum Matter Institute, University of British Columbia, Vancouver, BC V6T 1Z4, Canada}
\affiliation{Department of Physics \& Astronomy, University of British Columbia, Vancouver, BC V6T 1Z1, Canada}
\affiliation{Max Planck Institute for Solid State Research, Heisenbergstrasse 1, D-70569 Stuttgart, Germany}

\author{Yuta Murakami}
\affiliation{Department of Physics, Tokyo Institute of Technology, Meguro, Tokyo 152-8551, Japan}

\author{Giorgio Levy}
\affiliation{Quantum Matter Institute, University of British Columbia, Vancouver, BC V6T 1Z4, Canada}
\affiliation{Department of Physics \& Astronomy, University of British Columbia, Vancouver, BC V6T 1Z1, Canada}

\author{Arthur K. Mills}
\affiliation{Quantum Matter Institute, University of British Columbia, Vancouver, BC V6T 1Z4, Canada}
\affiliation{Department of Physics \& Astronomy, University of British Columbia, Vancouver, BC V6T 1Z1, Canada}

\author{Sergey Zhdanovich}
\affiliation{Quantum Matter Institute, University of British Columbia, Vancouver, BC V6T 1Z4, Canada}
\affiliation{Department of Physics \& Astronomy, University of British Columbia, Vancouver, BC V6T 1Z1, Canada}

\author{Masahiko Isobe}
\affiliation{Max Planck Institute for Solid State Research, Heisenbergstrasse 1, D-70569 Stuttgart, Germany}

\author{Hidenori Takagi}
\affiliation{Max Planck Institute for Solid State Research, Heisenbergstrasse 1, D-70569 Stuttgart, Germany}
\affiliation{Department of Physics, Tokyo Institute of Technology, Meguro, Tokyo 152-8551, Japan}

\author{Stefan Kaiser}
\affiliation{Max Planck Institute for Solid State Research, Heisenbergstrasse 1, D-70569 Stuttgart, Germany}

\author{Philipp Werner}
\affiliation{Department of Physics, University of Fribourg, 1700 Fribourg, Switzerland}

\author{David J. Jones}
\affiliation{Quantum Matter Institute, University of British Columbia, Vancouver, BC V6T 1Z4, Canada}
\affiliation{Department of Physics \& Astronomy, University of British Columbia, Vancouver, BC V6T 1Z1, Canada}

\author{Antoine Georges}
\affiliation{Coll\`ege de France, 11 place Marcelin Berthelot, 75005 Paris, France}
\affiliation{Center for Computational Quantum Physics, Flatiron Institute, 162 Fifth Avenue, New York, NY 10010, USA}
\affiliation{CPHT, CNRS, Ecole Polytechnique, IP Paris, F-91128 Palaiseau, France}
\affiliation{Department of Quantum Matter Physics, University of Geneva, 1211 Geneva 4, Switzerland}

\author{Andrea Damascelli}
\affiliation{Quantum Matter Institute, University of British Columbia, Vancouver, BC V6T 1Z4, Canada}
\affiliation{Department of Physics \& Astronomy, University of British Columbia, Vancouver, BC V6T 1Z1, Canada}

\author{Andrew J. Millis}
\affiliation{Center for Computational Quantum Physics, Flatiron Institute, 162 Fifth Avenue, New York, NY 10010, USA}
\affiliation{Department of Physics, Columbia University, 538 West 120th Street, New York, NY 10027}

\date{\today}

\begin{abstract}
We present a generic procedure for quantifying the interplay of electronic and lattice degrees of freedom in photo-doped insulators through a comparative analysis of theoretical many-body simulations and time- and angle-resolved photoemission spectroscopy~(TR-ARPES) of the transient response of the candidate excitonic insulator Ta$_2$NiSe$_5$. 
Our analysis demonstrates that the electron-electron interactions dominate the electron-phonon ones. In particular, a detailed analysis of the TR-ARPES spectrum enables a clear separation of the dominant broadening (electronic lifetime) effects from the much smaller bandgap renormalization. Theoretical calculations show that the observed strong spectral broadening arises from the electronic scattering of the photo-excited particle-hole pairs and cannot be accounted for in a model in which electron-phonon interactions are dominant. We demonstrate that the magnitude of the weaker subdominant bandgap renormalization sensitively depends on the distance from the semiconductor/semimetal transition in the high-temperature state, which could explain apparent contradictions between various TR-ARPES experiments. The analysis presented here indicates that electron-electron interactions play a vital role~(although not necessarily the sole one) in stabilizing the insulating state. 

\end{abstract}

\maketitle

The essence of strongly correlated electron physics
is understanding how novel ground states, such as high-temperature superconductivity, charge orders, and excitonic insulator~(EI) behavior as paradigmatic examples, emerge from competing degrees of freedom~\cite{keimer2017physics}. The excitonic insulator is a quantum many-body  state involving electron-hole pairing which can appear in semimetals or narrow gap semiconductors~\cite{keldysh1965,keldysh1968,des1965exciton,jerome1967,kohn1967}. In the semiconducting case, the excitonic binding energy exceeds the single particle gap leading to the Bose-Einstein condensation (BEC) of excitons. In the semimetallic case, the phase transition is described as a binding of weakly interacting electron-hole pairs~\cite{jerome1967,halperin1968excitonic} in analogy to the binding of electron-electron pairs described in the Bardeen-Cooper-Schrieffer~(BCS) theory of superconductivity. However, in solid state systems, an EI  is typically strongly coupled to lattice degrees of freedom and the interplay between electronic and electron-phonon interactions is a recurrent question in the field~\cite{kim2020,kimbj2021,volkov2020,ye2021,baldini2020,mor2017,mor2018}.

Recently, several systems have been proposed as excitonic insulator candidates, including 1T-TiSe$_2$~\cite{monney2012,monney2009}, Ta$_2$NiSe$_5$~(TNS)~\cite{lu2017zero,sunshine1985,wakisaka2009excitonic,he2021tunneling}, WTe$_2$~\cite{yanyu2020}, and Sb nanoflakes~\cite{li2019possible}. These materials have been the subject of extensive experimental exploration~\cite{monney2012,wang2019evidence,watson2020,werdehausen2018,kim2020,kim2020,volkov2020,ye2021} and theoretical modeling~\cite{zenker2014,kaneko2012,seki2014excitonic}, in an effort to provide definitive confirmation on the existence of the excitonic ground state and to elucidate the plethora of unique electronic and optical properties~\cite{butov2002,sun2020BaSh,nandi2012exciton,ma2021strongly}. TNS has  been of particular interest in this context. The characteristic flattening
of the valence band observed by angle-resolved photoemission spectroscopy~(ARPES)~\cite{seki2011,wakisaka2009excitonic} suggests a BCS pairing interpretation of the insulating state, while  the possibility of tuning the BCS-BEC transition via chemical or physical pressure~\cite{lu2017zero} opens new avenues for experimental investigation.

Early equilibrium ARPES results on TNS were interpreted within a purely electronic picture~\cite{seki2014excitonic,watson2020,mazza2020}, but subsequent additional experimental probes, including Raman~\cite{kim2020,kim2021,volkov2020,ye2021} and optical spectroscopy~\cite{werdehausen2018,larkin2017,larkin2018,andrich2020}, have suggested a strong coupling between the EI and lattice distortions raising the question of the quantitative
contribution of the electronic and lattice instabilities to the opening of the gap in TNS.
To resolve this question, 
TR-ARPES experiments have focused on the ultrafast response of the electronic gap, reporting either a transient modulation of the gap amplitude (interpreted within an electronic picture ~\cite{okazaki2018,tang2020,mor2017}) or a rigid gap response (interpreted within a largely lattice-driven scenario~\cite{baldini2020}).

Theoretical developments have followed a similar trajectory, where initial studies have focused on a purely excitonic description that successfully reproduced the equilibrium ARPES spectrum of TNS~\cite{seki2011,seki2014excitonic} and optical absorption spectra~\cite{sugimoto2018strong}. However, subsequent studies have revealed the importance of the electron-lattice coupling~\cite{subedi2020,baldini2020,windgatter2021}, and the interplay between electron-electron and electron-phonon contributions
is currently debated with nonlinear optical responses~\cite{werdehausen2018,golez2020,andrich2020}, collective dynamics~\cite{zenker2014,murakami2020}, and transient protocols for the order enhancement~\cite{murakami2017Photo,tanaka2018,tanabe2018,ryo2019} all discussed as highlighting either the electron-electron or electron-lattice contributions to the excitonic insulator state.

Motivated by these challenges, we performed a comparative analysis of photo-doped TNS using TR-ARPES and realistic model-system non-equilibrium many-body calculations that treat electronic and lattice degrees of freedom on equal footing. We show that the experimental response after photo-excitation is dominated by strong spectral broadening within the excitonic gap, which is well captured by microscopic simulations in the scenario of dominant 
electron-electron interactions.

\begin{figure}[t]
\includegraphics[width=0.9\linewidth]{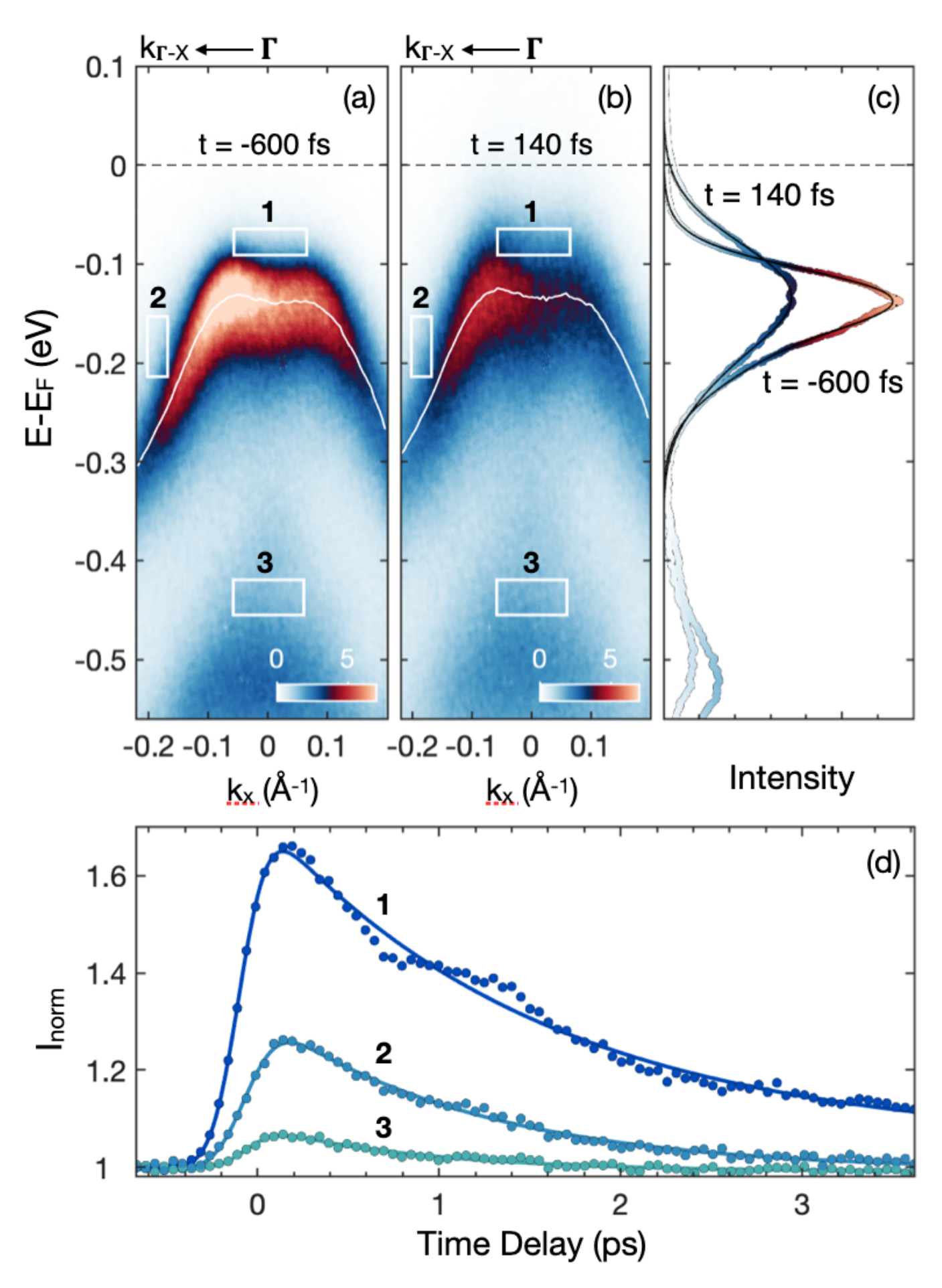}
\caption{TR-ARPES data
acquired along the $\Gamma$ - $X$ direction (a) before (t = -600 fs) (b) and after (t =  140 fs)  photoexcitation with a 50 $\mu$J/cm$^2$ incident pump fluence. (c) EDCs centered at $\Gamma$ before and after photoexcitation. (d) Time dependence of the integrated photoemission intensity over the three energy momentum regions outlined in white in panel (b) normalized to the equilibrium photoemission intensity in the corresponding regions. 
}
\label{Fig:exp1}
\end{figure}

TR-ARPES measurements were performed at the UBC-Moore Center for Ultrafast Quantum Matter by pumping with 1.55 eV photons (with incident fluence ranging from 26 to 160 muJ/cm2), and probing with 6.2 eV photons at a repetition rate of 250 kHz. The samples were aligned by Laue diffraction prior to the TR-ARPES measurements, and cleaved at 95 K in UHV at a base pressure better than 5$\times$10$^{-11}$ Torr. All TR-ARPES experiments were performed at a base cryostat temperature of 95 K; however, pump-induced thermal effects raising the sample's effective  base temperature have to be taken into account independently for each experiment. 
Both pump and probe beams were polarized parallel to the Ta-Ni chain direction, and the overall time and energy resolution of the system were 250 fs and 11 meV, respectively~\cite{Boshini2018}.

The equilibrium spectra [Fig.~\ref{Fig:exp1}(a)] show a characteristic ``M"-like flat upper valence band (UVB) centered at $\Gamma$, a dispersion that is characteristic of the EI ground state in the BCS regime~\cite{kaneko2014a,zenker2012}. Upon photoexcitation, we observe a significant decrease of the spectral weight. The response is clearly dominated  by the UVB's broadening and
to a lesser extent from a change in the peak position,
as shown by comparison of the photoemission spectra in panels (a) and (b) of Fig.~\ref{Fig:exp1}~(region 1 \& 2), and in the energy distribution curves (EDCs) presented in Fig.~\ref{Fig:exp1}(c). We quantify the two contributions
by extracting the full width at half maximum (FWHM) and the peak position $E_0$ from a fit of the UVB EDC spectral lineshape (the solid black line in Fig~\ref{Fig:exp1}(c) is the fit curve, details in \cite{SM}).

We analyze characteristic timescales of photo-induced effects by simple visualization of the energy- and momentum-integrated spectral intensity in selected areas as a function of pump-probe delay, see Fig.~\ref{Fig:exp1}(d). Each area is outside of the main band region, and all show a clear pump-induced increase in spectral intensity shortly after photoexcitation, followed by an exponential decay with a superimposed $\sim$1 THz oscillation observed in all areas. The latter is most apparent in region 1, approximately 1 ps after photoexcitation.

\begin{figure}[t]
\includegraphics[width=1.0\linewidth]{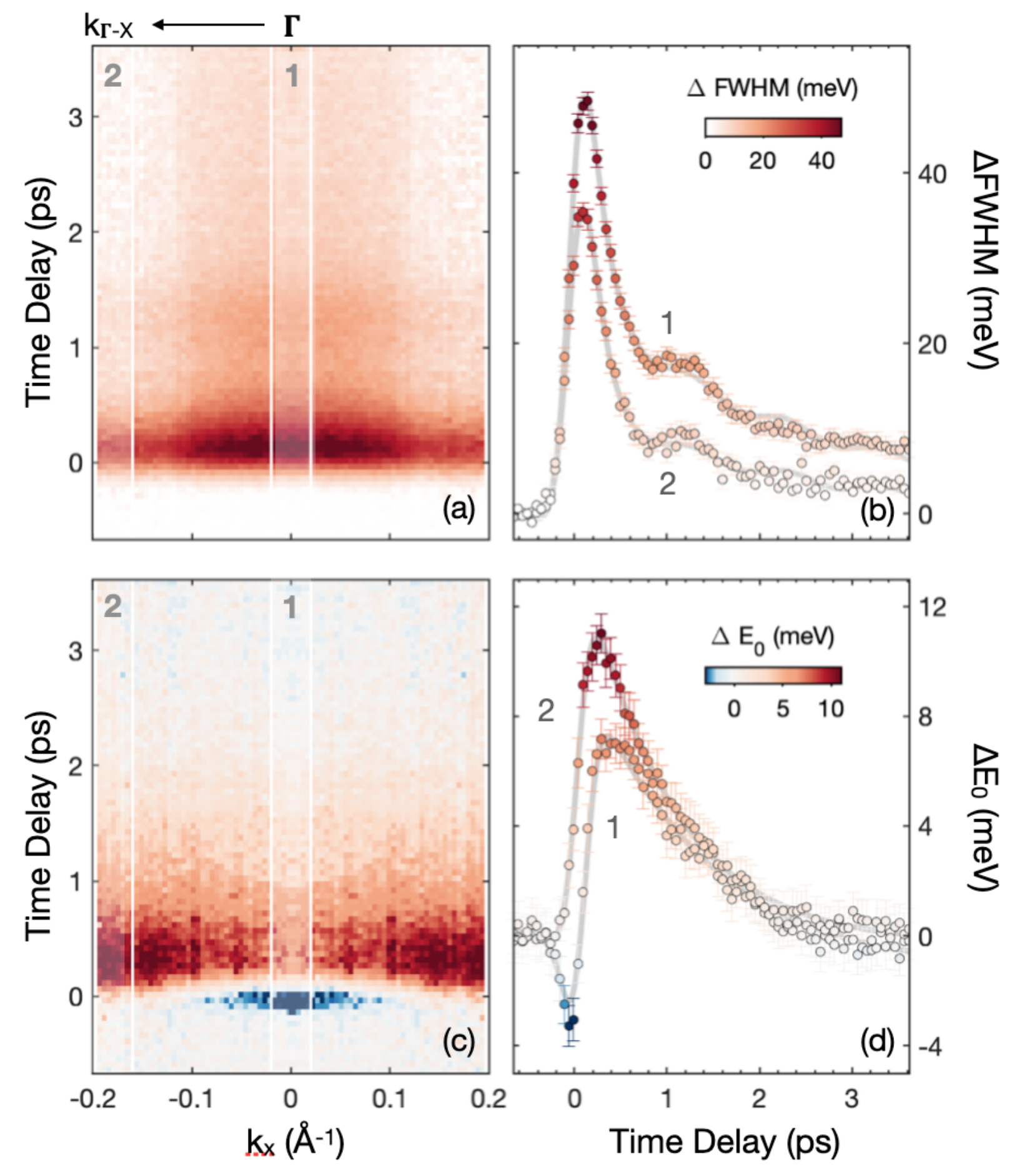}
\caption{(a) Momentum- and time-dependent change of the UVB width $\Delta$FWHM(k$_x$,t) for a 50 $\mu$J/cm$^2$ incident fluence. (b) One-dimensional traces of the UVB width as a function of pump-probe delay taken from select momentum integrated regions at $\Gamma$ (region 1), and at higher-momenta (region 2), outlined by the white shaded areas in (a). (c)-(d) Analog plots for momentum- and time-dependent change in the  UVB position $\Delta$E$_0$(k$_x$,t) with positive~(negative) values corresponding to the gap decrease~(increase).}
\label{Fig:exp2}
\end{figure} 
The time- and momentum-dependent $\Delta$FWHM of UVB is displayed in Fig.~\ref{Fig:exp2}(a), with selected traces around the $\Gamma$ point~(region 1), and at the highest momenta measured (region 2) in Fig.~\ref{Fig:exp2}(b). From Figs.~\ref{Fig:exp2}(a-b), one can observe a significant change in the broadening and relatively uniform rise-time across the entire k$_x$ region acquired, and subsequent relaxation with a superimposed $\sim$1 THz oscillation. The change in the gap amplitude [Figs.~\ref{Fig:exp2}(c-d)], in terms of a relative shift in the UVB position, shows a significantly different time-dependence than the linewidth. Unlike the width, the peak position shows no evidence of the $\sim$1 THz oscillation. This indicates that the oscillation observed in the normalized transient photoemission intensity [Fig.~\ref{Fig:exp1}(d), box 1 and 2] is a product of the UVB broadening and not an oscillation in the gap amplitude. Around $\Gamma$, the renormalization shows a non-monotonic response to photoexcitation, with a short-lived~($\sim$200 fs) resolution-limited gap-enhancement [downward shift in the UVB position to higher binding energies $\Delta$E$_0$(k$_x$,t)$<$0], which is consistent with Ref.~\onlinecite{mor2017}. The enhancement is followed by a partial gap-closure [upward relative shift in the UVB position to lower energies, $\Delta$E$_0$(k$_x$,t)$>$0], ending with an exponential recovery to equilibrium. On the contrary, we do not observe an ultrafast enhancement of the gap at higher momenta. The UVB shifts towards the Fermi level by a maximum of 11 meV (on the order of the energy resolution).  In summary, the pump-induced modifications to the photoemission spectrum are dominated by the UVB broadening, which is nearly an order of magnitude larger than the contribution of the bandgap renormalization. 

From here, we turn to theoretical modelling to qualitatively understand the microscopic implications of the observation. The model consists of two-band spinless fermions in one dimension coupled to dispersionless phonons:

\bsplit{
	&H=\sum_{k,\alpha\in\{0,1\}} [\epsilon_{k-A}]^{\alpha\alpha'} c^{\dagger}_{k,\alpha} c_{k,\alpha'}  + V \sum_{i} n_{i,0} n_{i,1} + \\
 	&\sum_i \left[\sqrt{\lambda} X_i -E(t)\right] c^{\dagger}_{k,0} c_{k,1} +\sum_i \frac{1}{2} [X_i^2+\frac{1}{\omega_0^2}\dot X_i^2]\text{+h.c.},
}
where $c_{k,\alpha}^{\dagger}$ is the electron creation operator for band $\alpha$ at momentum $k$, V is the Coulomb interaction strength between the conduction and valence band, $\lambda$ is the electron-phonon interaction strength corresponding to a displacement of $X = X_i$, and $E(A)$ is the electric field~(vector potential)~\cite{li2020,golez2019,dmytruk2020gauge}. The dispersion relation $\epsilon_{k}$ is obtained from the Wannier-interpolated DFT band structure in the high-temperature orthorhombic phase~\cite{mazza2020}. We mapped the Ta-Ni-Ta chain to the two-band problem by neglecting the interchain hopping as the smallest energy scale in the system~\cite{SM}.

Note that for all calculations presented, we include a weakly coupled bosonic bath not explicitly written in Eq. 1, so that the photo-induced electrons (holes) within maximum propagation time relax to the lower (upper) edge of the conduction (valence) band for all parameters.

\begin{figure}[t]
\includegraphics[width=0.8\linewidth]{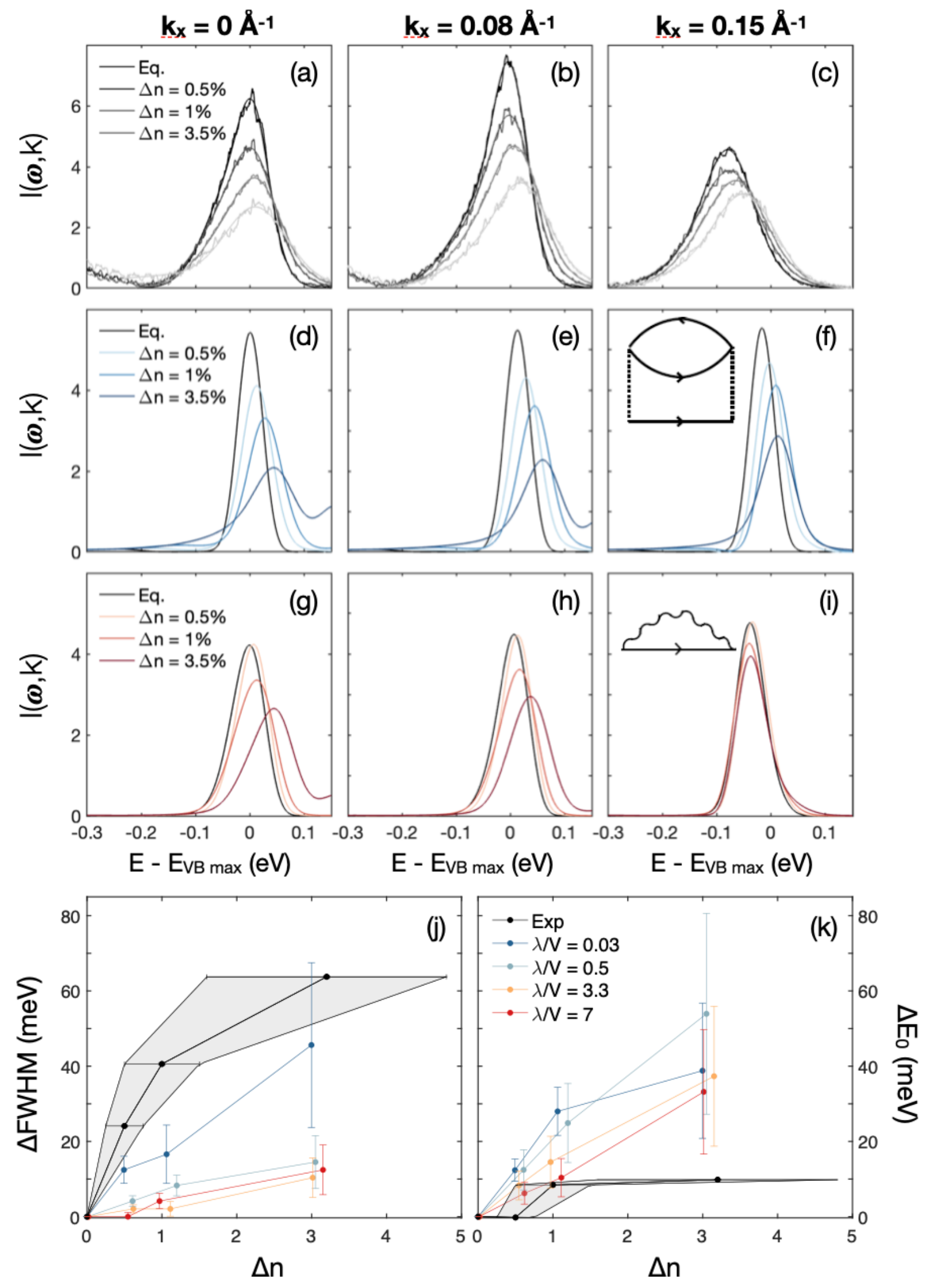}
\caption{Comparison of the photoemission spectrum in equilibrium~(Eq) and $t_p$=140 fs after the photo-excitation at three characteristic momenta along the $\Gamma$-X direction, namely k$_x$=0~(first column), k$_x$=0.08~A$^{-1}$~(second column) and k$_x$=0.15~A$^{-1}$~(third column), and for different excitation strengths~[$\Delta n=$ 0.5$\%$~(26 $\mu$J/cm$^2$), 1$\%$~(50 $\mu$J/cm$^2$), 3.5$\%$~(160 $\mu$J/cm$^2$)]. (a-c) Experimental EDCs; (d-f),(g-h) theoretical results in the electron~($V$=0.76 eV, $\lambda$=0.03 eV and $\omega_0$=0.016 eV) and the lattice~($V$=0.17 eV, $\lambda$=0.33 eV and $\omega_0$=0.016   eV) dominated regime, respectively. The inset presents the lowest-order scattering diagram for the two characteristic regimes. (j, k) The experimental and the theoretical change in width $\Delta$FWHM~(j) and position $\Delta$E$_0$~(k) of the UVB as a function of photo-doping $\Delta$n. Experimental errorbars represent uncertainty in the photo-doping estimation leading to theoretical uncertainties obtained from simulations at extreme values.}
\label{Fig:theo1}
\end{figure} 

The analysis of the photo-induced change in the lifetime calls for the inclusion of excitonic and lattice fluctuations on equal footing. We employ the Keldysh formalism with the 2$^{nd}$ Born~(Migdal) approximation for the electron-electron~(electron-lattice) scattering directly in the excitonic phase~\cite{SM}. First of all, we reproduce the experimental gap in equilibrium for various relative strengths of the electronic $V$ and electron-lattice interactions $\lambda$ at fixed gap size $2\Delta=0.25$ eV, and distinguish the electronically~($V\gg\lambda$) and lattice dominated~($V\ll\lambda$) case.  Afterwards, we estimate the relative importance of $V$ and $\lambda$ by analyzing photo-induced changes in the spectrum.

\begin{figure}[t]
\includegraphics[width=1.0\linewidth]{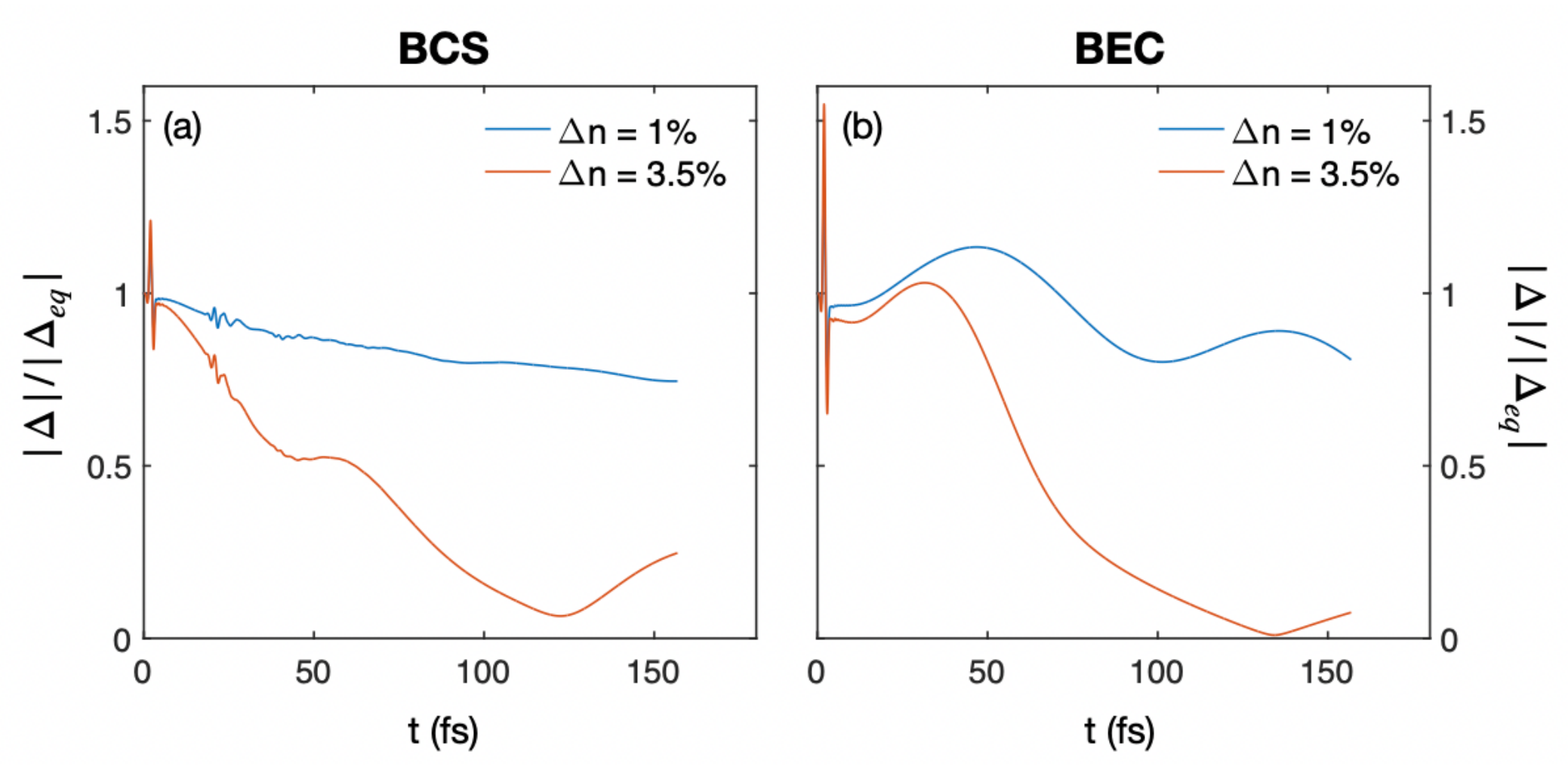}
\caption{Time-dependence of the order parameter in the BCS~(a) and the BEC~(b) regime after a photoexcitation for different excitation strengths.}
\label{Fig:theo2}
\end{figure} 

The system is excited by an electric field parametrized as $E(t)=A_0 \sin(\Omega t) \exp^{-4.2(t-t_0)^2},$ where $A_0$ is the pump strength, $\Omega$ is the 1.5 eV pump energy, an $t_0$ determines the single-cycle pump pulse. We adjust the pump strength $A_0$ such that the number of photo-excited electrons is fixed to $\Delta$n. 
The comparison between the theoretical~\cite{SM} and the experimentally determined photoemission 
EDCs~I$(\omega,k_x)$ for different momenta, before and after~(t$_p$=140 fs) photo-excitation are presented in Fig.~\ref{Fig:theo1}. Experimental data have been acquired at a repetition rate of 250 kHz, and steady-state heating of the sample has been observed and quantified for each fluence. The theoretical base temperature for higher fluences is adjusted by fixing the gap size to the experimental value before the pump pulse (details in \cite{SM}). The theoretically derived line profiles were determined for two characteristic regimes: (i) the primarily electron-driven case [$V$ = 0.92 eV, $\lambda$ = 0.03 eV], and (ii) the primarily lattice-driven case [$V$ = 0.13 eV, $\lambda$ = 0.43 eV].

In equilibrium, the width of the UVB is consistently larger in the experimental data [black curves in Fig.~\ref{Fig:theo1}(a-c)]  than in the theoretical predictions. This is likely to originate from the presence of additional scattering channels (different phonons, impurities, etc.). By simple visual comparison of the non-equilibrium results in Fig.~\ref{Fig:theo1}, we note that the significant broadening of the UVB is only captured in the primarily electron-driven case [Fig.~\ref{Fig:theo1}(d-f)]. In contrast, the transient variation of the width in the lattice-driven case is substantially smaller.

We compare the theoretical and experimental results more quantitatively by extracting the photo-induced change in the width ($\Delta$FWHM), proportional to the quasiparticle lifetime change, and the peak position ($\Delta $E$_0$) of the UVB EDC centered at $\Gamma$ as a function of photodoping $\Delta$n~\cite{SM}. We note that, although experimental and theoretical $\Delta$FWHM display different absolute variations, the trend as a function of the photoexcitation  density is  better captured by the pure electronic-driven case, see Fig.~\ref{Fig:theo1}(j) (blue curve). Note particularly that in the primarily phonon-driven case, the linewidth displays only a weak dependence on the photodoping excitation, in sharp contrast to experiment.

An analogous comparison for the subdominant bandgap renormalization $\Delta$E$_0$ is presented in Fig.~\ref{Fig:theo1}(k). The photo-induced modification in  the band position is smaller than the lifetime change. There is an apparent saturation at larger photo-doping ($\Delta$n$>$ 0.01) in the experimental data [black data, Fig.~\ref{Fig:theo1}(k)], which notably occurs below our system energy resolution of 11 meV. Although the experimental and theoretical shifts agree for the smallest photo-doping considered~($\Delta n\leq 1 \%$), the theoretically derived results do not saturate and lead to larger bandgap renormalization.

We now discuss the possible microscopic origins of these observations. The photo-excitation introduces mobile charge carriers into the conduction and valence bands, changing effectively the system's scatterings. The change in the electronic scattering is proportional to the number of photo-doped charge carriers which modifies the particle-hole bubble in the lowest order scattering diagram, see inset of Fig.~\ref{Fig:theo1}(f). On the contrary, the lowest order electron-phonon scattering, see inset of Fig.~\ref{Fig:theo1}(i), can be estimated from the Fermi golden rule as $\Sigma(\omega)\propto \lambda \rho(\omega-\omega_0),$ where neither the electronic  density of states $\rho(\omega)$ nor $\lambda$ has direct photo-doping dependence, resulting in an electron lattice scattering rate with weak dependence on the photo-excitation intensity. While the self-consistent Migdal approximation allows for the dressing of the lattice propagator, our numerical results show that the  broadening is small and primarily originates from heating effects due to the finite repetition rate. Therefore, the qualitative difference in the lifetime scaling can be ascribed to the distinct scatterings between the electronic and lattice degrees of freedom. The qualitative agreement between the experimental and theoretical photodoping dependence of the lifetime suggests a robust electronic character in TNS.

Finally, the experimental bandgap renormalization close to the $\Gamma$ point is, surprisingly, small;  importantly, we even observed a transient enhancement [see Fig.~\ref{Fig:exp2}(c)]. We investigated the gap size evolution after a nonlinear excitation, and we show that it is highly susceptible to the position in the BCS-BEC crossover~\cite{zenker2012}, as revealed by the dynamics of the excitonic order parameter $\phi=\sum_k \langle c_{1,k}^{\dagger} c_{0,k} \rangle$ being a direct measure of the gap size. In the (semimetallic)~BCS regime, we report a monotonic reduction of the order parameter upon optical excitation, see Fig.~\ref{Fig:theo2}(a). On the contrary, a transient enhancement of order parameter can be observed in the~(semiconducting) BEC regime [Fig.~\ref{Fig:theo2}(b)], in agreement with the experimental findings [Fig.~\ref{Fig:exp2}(b)]. 
The enhancement is only transient due to the inclusion of scatterings, in contrast to previous time-dependent mean-field studies predicting a long-lasting enhancement~\cite{murakami2017Photo,tanaka2018,tanabe2018,ryo2019}. Implications of these results are twofold. First, the sensitivity of the gap renormalization dynamics to the position in the BEC-BCS crossover may explain the apparent contradiction between theory and experiment in Fig.~\ref{Fig:theo1}(k), as well as various TR-ARPES experiments which showed a wide range of responses, including gap reductions~\cite{okazaki2018,tang2020,saha2021}, gap enhancements~\cite{mor2017} and rigid gap shifts~\cite{baldini2020}. Second, the contradiction between equilibrium ARPES results, suggesting a BCS nature of the ground state, and the transient gap enhancement observed in TRARPES remains, an open question.

In conclusion, we have reported a marked photo-induced broadening of the valence band of TNS, while the bandgap renormalization is almost an order of magnitude smaller. A detailed comparison of the TR-ARPES experimental data with nonequilibrium many-body simulations demonstrates the pivotal~(but not necessarily sole) contribution of electron-electron interactions to the stabilization of the excitonic gap in TNS.

More broadly, this work demonstrates a generic way to determine the origin of the gap in correlated insulators by analyzing the lifetime effects of photo-doped states. Furthermore, the bandgap enhancement after a nonlinear excitation shows a subtle dependence on the position in the BCS-BEC crossover. In this regard, we propose a systematic study of the nonlinear response by applying either chemical or physical pressure, as performed in recent transport~\cite{lu2017zero,matsubayashi2021hybridization} and Raman measurements~\cite{ye2021}, to optimize the nonthermal enhancement of the underlying order.

\begin{acknowledgments}

DG is supported by Slovenian Research Agency (ARRS) under Program J1-2455 and P1-0044. The Flatiron Institute is a division of the Simons Foundation. A. J. M. acknowledges support from the Energy Frontier Research Center on Programmable Quantum Materials funded by the US Department of Energy (DOE), Office of Science, Basic Energy Sciences (BES), under Award No. de-sc0019443. A. G. acknowledges support from the European Research Council (ERCQMAC-319286). P.W. acknowledges support from ERC Consolidator Grant No. 724103. Y.M. is supported by Grant-in-Aid for Scientific Research from JSPS, KAKENHI Grant Nos. JP20K14412 and JP21H05017. This research was undertaken thanks, in part, to funding from the Max Planck-UBC-UTokyo Center for Quantum Materials and the Canada First Research Excellence Fund, Quantum Materials, and Future Technologies Program. This project is also funded by the Gordon and Betty Moore Foundation’s EPiQS Initiative, Grant No. GBMF4779 to A.D. and D.J.J.; the Natural Sciences and Engineering Research Council of Canada (NSERC); the Canada Foundation for Innovation (CFI); the British Columbia Knowledge Development Fund (BCKDF); the Canada Research Chairs Program (A.D.); and the CIFAR Quantum Materials Program (A.D.). We numerically solved the time-dependent problem using the library NESSi~\cite{schuler2020nessi}. D.G. acknowledges fruitful discussions with Tatsuya Kaneko and Zhiyuan Sun.
\end{acknowledgments}

\bibliography{tdmft.bib,excitons.bib,Books}

\begin{thebibliography}{62}%
\makeatletter
\providecommand \@ifxundefined [1]{%
 \@ifx{#1\undefined}
}%
\providecommand \@ifnum [1]{%
 \ifnum #1\expandafter \@firstoftwo
 \else \expandafter \@secondoftwo
 \fi
}%
\providecommand \@ifx [1]{%
 \ifx #1\expandafter \@firstoftwo
 \else \expandafter \@secondoftwo
 \fi
}%
\providecommand \natexlab [1]{#1}%
\providecommand \enquote  [1]{``#1''}%
\providecommand \bibnamefont  [1]{#1}%
\providecommand \bibfnamefont [1]{#1}%
\providecommand \citenamefont [1]{#1}%
\providecommand \href@noop [0]{\@secondoftwo}%
\providecommand \href [0]{\begingroup \@sanitize@url \@href}%
\providecommand \@href[1]{\@@startlink{#1}\@@href}%
\providecommand \@@href[1]{\endgroup#1\@@endlink}%
\providecommand \@sanitize@url [0]{\catcode `\\12\catcode `\$12\catcode
  `\&12\catcode `\#12\catcode `\^12\catcode `\_12\catcode `\%12\relax}%
\providecommand \@@startlink[1]{}%
\providecommand \@@endlink[0]{}%
\providecommand \url  [0]{\begingroup\@sanitize@url \@url }%
\providecommand \@url [1]{\endgroup\@href {#1}{\urlprefix }}%
\providecommand \urlprefix  [0]{URL }%
\providecommand \Eprint [0]{\href }%
\providecommand \doibase [0]{http://dx.doi.org/}%
\providecommand \selectlanguage [0]{\@gobble}%
\providecommand \bibinfo  [0]{\@secondoftwo}%
\providecommand \bibfield  [0]{\@secondoftwo}%
\providecommand \translation [1]{[#1]}%
\providecommand \BibitemOpen [0]{}%
\providecommand \bibitemStop [0]{}%
\providecommand \bibitemNoStop [0]{.\EOS\space}%
\providecommand \EOS [0]{\spacefactor3000\relax}%
\providecommand \BibitemShut  [1]{\csname bibitem#1\endcsname}%
\let\auto@bib@innerbib\@empty
%</preamble>
\bibitem [{\citenamefont {Keimer}\ and\ \citenamefont
  {Moore}(2017)}]{keimer2017physics}%
  \BibitemOpen
  \bibfield  {author} {\bibinfo {author} {\bibfnamefont {B.}~\bibnamefont
  {Keimer}}\ and\ \bibinfo {author} {\bibfnamefont {J.}~\bibnamefont {Moore}},\
  }\href@noop {} {\bibfield  {journal} {\bibinfo  {journal} {Nature Physics}\
  }\textbf {\bibinfo {volume} {13}},\ \bibinfo {pages} {1045} (\bibinfo {year}
  {2017})}\BibitemShut {NoStop}%
\bibitem [{\citenamefont {Keldysh}\ and\ \citenamefont
  {Kopaev}(1965)}]{keldysh1965}%
  \BibitemOpen
  \bibfield  {author} {\bibinfo {author} {\bibfnamefont {L.}~\bibnamefont
  {Keldysh}}\ and\ \bibinfo {author} {\bibfnamefont {Y.~V.}\ \bibnamefont
  {Kopaev}},\ }\href@noop {} {\bibfield  {journal} {\bibinfo  {journal} {Soviet
  Physics Solid State, USSR}\ }\textbf {\bibinfo {volume} {6}},\ \bibinfo
  {pages} {2219} (\bibinfo {year} {1965})}\BibitemShut {NoStop}%
\bibitem [{\citenamefont {{Keldysh}}\ and\ \citenamefont
  {{Kozlov}}(1968)}]{keldysh1968}%
  \BibitemOpen
  \bibfield  {author} {\bibinfo {author} {\bibfnamefont {L.~V.}\ \bibnamefont
  {{Keldysh}}}\ and\ \bibinfo {author} {\bibfnamefont {A.~N.}\ \bibnamefont
  {{Kozlov}}},\ }\href@noop {} {\bibfield  {journal} {\bibinfo  {journal}
  {Soviet Journal of Experimental and Theoretical Physics}\ }\textbf {\bibinfo
  {volume} {27}},\ \bibinfo {pages} {521} (\bibinfo {year} {1968})}\BibitemShut
  {NoStop}%
\bibitem [{\citenamefont {Des~Cloizeaux}(1965)}]{des1965exciton}%
  \BibitemOpen
  \bibfield  {author} {\bibinfo {author} {\bibfnamefont {J.}~\bibnamefont
  {Des~Cloizeaux}},\ }\href@noop {} {\bibfield  {journal} {\bibinfo  {journal}
  {Journal of Physics and Chemistry of Solids}\ }\textbf {\bibinfo {volume}
  {26}},\ \bibinfo {pages} {259} (\bibinfo {year} {1965})}\BibitemShut
  {NoStop}%
\bibitem [{\citenamefont {J\'erome}\ \emph {et~al.}(1967)\citenamefont
  {J\'erome}, \citenamefont {Rice},\ and\ \citenamefont {Kohn}}]{jerome1967}%
  \BibitemOpen
  \bibfield  {author} {\bibinfo {author} {\bibfnamefont {D.}~\bibnamefont
  {J\'erome}}, \bibinfo {author} {\bibfnamefont {T.~M.}\ \bibnamefont {Rice}},
  \ and\ \bibinfo {author} {\bibfnamefont {W.}~\bibnamefont {Kohn}},\ }\href
  {\doibase 10.1103/PhysRev.158.462} {\bibfield  {journal} {\bibinfo  {journal}
  {Phys. Rev.}\ }\textbf {\bibinfo {volume} {158}},\ \bibinfo {pages} {462}
  (\bibinfo {year} {1967})}\BibitemShut {NoStop}%
\bibitem [{\citenamefont {Kohn}(1967)}]{kohn1967}%
  \BibitemOpen
  \bibfield  {author} {\bibinfo {author} {\bibfnamefont {W.}~\bibnamefont
  {Kohn}},\ }\href {\doibase 10.1103/PhysRevLett.19.439} {\bibfield  {journal}
  {\bibinfo  {journal} {Phys. Rev. Lett.}\ }\textbf {\bibinfo {volume} {19}},\
  \bibinfo {pages} {439} (\bibinfo {year} {1967})}\BibitemShut {NoStop}%
\bibitem [{\citenamefont {Halperin}\ and\ \citenamefont
  {Rice}(1968)}]{halperin1968excitonic}%
  \BibitemOpen
  \bibfield  {author} {\bibinfo {author} {\bibfnamefont {B.}~\bibnamefont
  {Halperin}}\ and\ \bibinfo {author} {\bibfnamefont {T.}~\bibnamefont
  {Rice}},\ }in\ \href@noop {} {\emph {\bibinfo {booktitle} {Solid State
  Physics}}},\ Vol.~\bibinfo {volume} {21}\ (\bibinfo  {publisher} {Elsevier},\
  \bibinfo {year} {1968})\ pp.\ \bibinfo {pages} {115--192}\BibitemShut
  {NoStop}%
\bibitem [{\citenamefont {Kim}\ \emph {et~al.}(2020)\citenamefont {Kim},
  \citenamefont {Schulz}, \citenamefont {Takayama}, \citenamefont {Isobe},
  \citenamefont {Takagi},\ and\ \citenamefont {Kaiser}}]{kim2020}%
  \BibitemOpen
  \bibfield  {author} {\bibinfo {author} {\bibfnamefont {M.-J.}\ \bibnamefont
  {Kim}}, \bibinfo {author} {\bibfnamefont {A.}~\bibnamefont {Schulz}},
  \bibinfo {author} {\bibfnamefont {T.}~\bibnamefont {Takayama}}, \bibinfo
  {author} {\bibfnamefont {M.}~\bibnamefont {Isobe}}, \bibinfo {author}
  {\bibfnamefont {H.}~\bibnamefont {Takagi}}, \ and\ \bibinfo {author}
  {\bibfnamefont {S.}~\bibnamefont {Kaiser}},\ }\href {\doibase
  10.1103/PhysRevResearch.2.042039} {\bibfield  {journal} {\bibinfo  {journal}
  {Phys. Rev. Research}\ }\textbf {\bibinfo {volume} {2}},\ \bibinfo {pages}
  {042039} (\bibinfo {year} {2020})}\BibitemShut {NoStop}%
\bibitem [{\citenamefont {Kim}\ \emph {et~al.}(2021{\natexlab{a}})\citenamefont
  {Kim}, \citenamefont {Kim}, \citenamefont {Kim}, \citenamefont {Kwon},
  \citenamefont {Kim},\ and\ \citenamefont {Kim}}]{kimbj2021}%
  \BibitemOpen
  \bibfield  {author} {\bibinfo {author} {\bibfnamefont {K.}~\bibnamefont
  {Kim}}, \bibinfo {author} {\bibfnamefont {H.}~\bibnamefont {Kim}}, \bibinfo
  {author} {\bibfnamefont {J.}~\bibnamefont {Kim}}, \bibinfo {author}
  {\bibfnamefont {C.}~\bibnamefont {Kwon}}, \bibinfo {author} {\bibfnamefont
  {J.~S.}\ \bibnamefont {Kim}}, \ and\ \bibinfo {author} {\bibfnamefont
  {B.}~\bibnamefont {Kim}},\ }\href@noop {} {\bibfield  {journal} {\bibinfo
  {journal} {Nature communications}\ }\textbf {\bibinfo {volume} {12}},\
  \bibinfo {pages} {1} (\bibinfo {year} {2021}{\natexlab{a}})}\BibitemShut
  {NoStop}%
\bibitem [{\citenamefont {Volkov}\ \emph {et~al.}(2021)\citenamefont {Volkov},
  \citenamefont {Ye}, \citenamefont {Lohani}, \citenamefont {Feldman},
  \citenamefont {Kanigel},\ and\ \citenamefont {Blumberg}}]{volkov2020}%
  \BibitemOpen
  \bibfield  {author} {\bibinfo {author} {\bibfnamefont {P.}~\bibnamefont
  {Volkov}}, \bibinfo {author} {\bibfnamefont {M.}~\bibnamefont {Ye}}, \bibinfo
  {author} {\bibfnamefont {H.}~\bibnamefont {Lohani}}, \bibinfo {author}
  {\bibfnamefont {I.}~\bibnamefont {Feldman}}, \bibinfo {author} {\bibfnamefont
  {A.}~\bibnamefont {Kanigel}}, \ and\ \bibinfo {author} {\bibfnamefont
  {G.}~\bibnamefont {Blumberg}},\ }\href@noop {} {\bibfield  {journal}
  {\bibinfo  {journal} {npj Quantum Materials}\ }\textbf {\bibinfo {volume}
  {6}},\ \bibinfo {pages} {1} (\bibinfo {year} {2021})}\BibitemShut {NoStop}%
\bibitem [{\citenamefont {Ye}\ \emph {et~al.}(2021)\citenamefont {Ye},
  \citenamefont {Volkov}, \citenamefont {Lohani}, \citenamefont {Feldman},
  \citenamefont {Kim}, \citenamefont {Kanigel},\ and\ \citenamefont
  {Blumberg}}]{ye2021}%
  \BibitemOpen
  \bibfield  {author} {\bibinfo {author} {\bibfnamefont {M.}~\bibnamefont
  {Ye}}, \bibinfo {author} {\bibfnamefont {P.~A.}\ \bibnamefont {Volkov}},
  \bibinfo {author} {\bibfnamefont {H.}~\bibnamefont {Lohani}}, \bibinfo
  {author} {\bibfnamefont {I.}~\bibnamefont {Feldman}}, \bibinfo {author}
  {\bibfnamefont {M.}~\bibnamefont {Kim}}, \bibinfo {author} {\bibfnamefont
  {A.}~\bibnamefont {Kanigel}}, \ and\ \bibinfo {author} {\bibfnamefont
  {G.}~\bibnamefont {Blumberg}},\ }\href@noop {} {\bibfield  {journal}
  {\bibinfo  {journal} {Physical Review B}\ }\textbf {\bibinfo {volume}
  {104}},\ \bibinfo {pages} {045102} (\bibinfo {year} {2021})}\BibitemShut
  {NoStop}%
\bibitem [{\citenamefont {{Baldini}}\ \emph {et~al.}(2020)\citenamefont
  {{Baldini}}, \citenamefont {{Zong}}, \citenamefont {{Choi}}, \citenamefont
  {{Lee}}, \citenamefont {{Michael}}, \citenamefont {{Windgaetter}},
  \citenamefont {{Mazin}}, \citenamefont {{Latini}}, \citenamefont {{Azoury}},
  \citenamefont {{Lv}}, \citenamefont {{Kogar}}, \citenamefont {{Wang}},
  \citenamefont {{Lu}}, \citenamefont {{Takayama}}, \citenamefont {{Takagi}},
  \citenamefont {{Millis}}, \citenamefont {{Rubio}}, \citenamefont {{Demler}},\
  and\ \citenamefont {{Gedik}}}]{baldini2020}%
  \BibitemOpen
  \bibfield  {author} {\bibinfo {author} {\bibfnamefont {E.}~\bibnamefont
  {{Baldini}}}, \bibinfo {author} {\bibfnamefont {A.}~\bibnamefont {{Zong}}},
  \bibinfo {author} {\bibfnamefont {D.}~\bibnamefont {{Choi}}}, \bibinfo
  {author} {\bibfnamefont {C.}~\bibnamefont {{Lee}}}, \bibinfo {author}
  {\bibfnamefont {M.~H.}\ \bibnamefont {{Michael}}}, \bibinfo {author}
  {\bibfnamefont {L.}~\bibnamefont {{Windgaetter}}}, \bibinfo {author}
  {\bibfnamefont {I.~I.}\ \bibnamefont {{Mazin}}}, \bibinfo {author}
  {\bibfnamefont {S.}~\bibnamefont {{Latini}}}, \bibinfo {author}
  {\bibfnamefont {D.}~\bibnamefont {{Azoury}}}, \bibinfo {author}
  {\bibfnamefont {B.}~\bibnamefont {{Lv}}}, \bibinfo {author} {\bibfnamefont
  {A.}~\bibnamefont {{Kogar}}}, \bibinfo {author} {\bibfnamefont
  {Y.}~\bibnamefont {{Wang}}}, \bibinfo {author} {\bibfnamefont
  {Y.}~\bibnamefont {{Lu}}}, \bibinfo {author} {\bibfnamefont {T.}~\bibnamefont
  {{Takayama}}}, \bibinfo {author} {\bibfnamefont {H.}~\bibnamefont
  {{Takagi}}}, \bibinfo {author} {\bibfnamefont {A.~J.}\ \bibnamefont
  {{Millis}}}, \bibinfo {author} {\bibfnamefont {A.}~\bibnamefont {{Rubio}}},
  \bibinfo {author} {\bibfnamefont {E.}~\bibnamefont {{Demler}}}, \ and\
  \bibinfo {author} {\bibfnamefont {N.}~\bibnamefont {{Gedik}}},\ }\href@noop
  {} {\bibfield  {journal} {\bibinfo  {journal} {arXiv e-prints}\ ,\ \bibinfo
  {eid} {arXiv:2007.02909}} (\bibinfo {year} {2020})},\ \Eprint
  {http://arxiv.org/abs/2007.02909} {arXiv:2007.02909 [cond-mat.str-el]}
  \BibitemShut {NoStop}%
\bibitem [{\citenamefont {Mor}\ \emph {et~al.}(2017)\citenamefont {Mor},
  \citenamefont {Herzog}, \citenamefont {Gole\ifmmode~\check{z}\else
  \v{z}\fi{}}, \citenamefont {Werner}, \citenamefont {Eckstein}, \citenamefont
  {Katayama}, \citenamefont {Nohara}, \citenamefont {Takagi}, \citenamefont
  {Mizokawa}, \citenamefont {Monney},\ and\ \citenamefont
  {St\"ahler}}]{mor2017}%
  \BibitemOpen
  \bibfield  {author} {\bibinfo {author} {\bibfnamefont {S.}~\bibnamefont
  {Mor}}, \bibinfo {author} {\bibfnamefont {M.}~\bibnamefont {Herzog}},
  \bibinfo {author} {\bibfnamefont {D.}~\bibnamefont
  {Gole\ifmmode~\check{z}\else \v{z}\fi{}}}, \bibinfo {author} {\bibfnamefont
  {P.}~\bibnamefont {Werner}}, \bibinfo {author} {\bibfnamefont
  {M.}~\bibnamefont {Eckstein}}, \bibinfo {author} {\bibfnamefont
  {N.}~\bibnamefont {Katayama}}, \bibinfo {author} {\bibfnamefont
  {M.}~\bibnamefont {Nohara}}, \bibinfo {author} {\bibfnamefont
  {H.}~\bibnamefont {Takagi}}, \bibinfo {author} {\bibfnamefont
  {T.}~\bibnamefont {Mizokawa}}, \bibinfo {author} {\bibfnamefont
  {C.}~\bibnamefont {Monney}}, \ and\ \bibinfo {author} {\bibfnamefont
  {J.}~\bibnamefont {St\"ahler}},\ }\href {\doibase
  10.1103/PhysRevLett.119.086401} {\bibfield  {journal} {\bibinfo  {journal}
  {Phys. Rev. Lett.}\ }\textbf {\bibinfo {volume} {119}},\ \bibinfo {pages}
  {086401} (\bibinfo {year} {2017})}\BibitemShut {NoStop}%
\bibitem [{\citenamefont {Mor}\ \emph {et~al.}(2018)\citenamefont {Mor},
  \citenamefont {Herzog}, \citenamefont {Noack}, \citenamefont {Katayama},
  \citenamefont {Nohara}, \citenamefont {Takagi}, \citenamefont {Trunschke},
  \citenamefont {Mizokawa}, \citenamefont {Monney},\ and\ \citenamefont
  {St\"ahler}}]{mor2018}%
  \BibitemOpen
  \bibfield  {author} {\bibinfo {author} {\bibfnamefont {S.}~\bibnamefont
  {Mor}}, \bibinfo {author} {\bibfnamefont {M.}~\bibnamefont {Herzog}},
  \bibinfo {author} {\bibfnamefont {J.}~\bibnamefont {Noack}}, \bibinfo
  {author} {\bibfnamefont {N.}~\bibnamefont {Katayama}}, \bibinfo {author}
  {\bibfnamefont {M.}~\bibnamefont {Nohara}}, \bibinfo {author} {\bibfnamefont
  {H.}~\bibnamefont {Takagi}}, \bibinfo {author} {\bibfnamefont
  {A.}~\bibnamefont {Trunschke}}, \bibinfo {author} {\bibfnamefont
  {T.}~\bibnamefont {Mizokawa}}, \bibinfo {author} {\bibfnamefont
  {C.}~\bibnamefont {Monney}}, \ and\ \bibinfo {author} {\bibfnamefont
  {J.}~\bibnamefont {St\"ahler}},\ }\href {\doibase 10.1103/PhysRevB.97.115154}
  {\bibfield  {journal} {\bibinfo  {journal} {Phys. Rev. B}\ }\textbf {\bibinfo
  {volume} {97}},\ \bibinfo {pages} {115154} (\bibinfo {year}
  {2018})}\BibitemShut {NoStop}%
\bibitem [{\citenamefont {Monney}\ \emph {et~al.}(2012)\citenamefont {Monney},
  \citenamefont {Monney}, \citenamefont {Aebi},\ and\ \citenamefont
  {Beck}}]{monney2012}%
  \BibitemOpen
  \bibfield  {author} {\bibinfo {author} {\bibfnamefont {C.}~\bibnamefont
  {Monney}}, \bibinfo {author} {\bibfnamefont {G.}~\bibnamefont {Monney}},
  \bibinfo {author} {\bibfnamefont {P.}~\bibnamefont {Aebi}}, \ and\ \bibinfo
  {author} {\bibfnamefont {H.}~\bibnamefont {Beck}},\ }\href@noop {} {\bibfield
   {journal} {\bibinfo  {journal} {New Journal of Physics}\ }\textbf {\bibinfo
  {volume} {14}},\ \bibinfo {pages} {075026} (\bibinfo {year}
  {2012})}\BibitemShut {NoStop}%
\bibitem [{\citenamefont {Monney}\ \emph {et~al.}(2009)\citenamefont {Monney},
  \citenamefont {Cercellier}, \citenamefont {Clerc}, \citenamefont {Battaglia},
  \citenamefont {Schwier}, \citenamefont {Didiot}, \citenamefont {Garnier},
  \citenamefont {Beck}, \citenamefont {Aebi}, \citenamefont {Berger},
  \citenamefont {Forr\'o},\ and\ \citenamefont {Patthey}}]{monney2009}%
  \BibitemOpen
  \bibfield  {author} {\bibinfo {author} {\bibfnamefont {C.}~\bibnamefont
  {Monney}}, \bibinfo {author} {\bibfnamefont {H.}~\bibnamefont {Cercellier}},
  \bibinfo {author} {\bibfnamefont {F.}~\bibnamefont {Clerc}}, \bibinfo
  {author} {\bibfnamefont {C.}~\bibnamefont {Battaglia}}, \bibinfo {author}
  {\bibfnamefont {E.~F.}\ \bibnamefont {Schwier}}, \bibinfo {author}
  {\bibfnamefont {C.}~\bibnamefont {Didiot}}, \bibinfo {author} {\bibfnamefont
  {M.~G.}\ \bibnamefont {Garnier}}, \bibinfo {author} {\bibfnamefont
  {H.}~\bibnamefont {Beck}}, \bibinfo {author} {\bibfnamefont {P.}~\bibnamefont
  {Aebi}}, \bibinfo {author} {\bibfnamefont {H.}~\bibnamefont {Berger}},
  \bibinfo {author} {\bibfnamefont {L.}~\bibnamefont {Forr\'o}}, \ and\
  \bibinfo {author} {\bibfnamefont {L.}~\bibnamefont {Patthey}},\ }\href
  {\doibase 10.1103/PhysRevB.79.045116} {\bibfield  {journal} {\bibinfo
  {journal} {Phys. Rev. B}\ }\textbf {\bibinfo {volume} {79}},\ \bibinfo
  {pages} {045116} (\bibinfo {year} {2009})}\BibitemShut {NoStop}%
\bibitem [{\citenamefont {Lu}\ \emph {et~al.}(2017)\citenamefont {Lu},
  \citenamefont {Kono}, \citenamefont {Larkin}, \citenamefont {Rost},
  \citenamefont {Takayama}, \citenamefont {Boris}, \citenamefont {Keimer},\
  and\ \citenamefont {Takagi}}]{lu2017zero}%
  \BibitemOpen
  \bibfield  {author} {\bibinfo {author} {\bibfnamefont {Y.}~\bibnamefont
  {Lu}}, \bibinfo {author} {\bibfnamefont {H.}~\bibnamefont {Kono}}, \bibinfo
  {author} {\bibfnamefont {T.}~\bibnamefont {Larkin}}, \bibinfo {author}
  {\bibfnamefont {A.}~\bibnamefont {Rost}}, \bibinfo {author} {\bibfnamefont
  {T.}~\bibnamefont {Takayama}}, \bibinfo {author} {\bibfnamefont
  {A.}~\bibnamefont {Boris}}, \bibinfo {author} {\bibfnamefont
  {B.}~\bibnamefont {Keimer}}, \ and\ \bibinfo {author} {\bibfnamefont
  {H.}~\bibnamefont {Takagi}},\ }\href@noop {} {\bibfield  {journal} {\bibinfo
  {journal} {Nature communications}\ }\textbf {\bibinfo {volume} {8}},\
  \bibinfo {pages} {14408} (\bibinfo {year} {2017})}\BibitemShut {NoStop}%
\bibitem [{\citenamefont {Sunshine}\ and\ \citenamefont
  {Ibers}(1985)}]{sunshine1985}%
  \BibitemOpen
  \bibfield  {author} {\bibinfo {author} {\bibfnamefont {S.~A.}\ \bibnamefont
  {Sunshine}}\ and\ \bibinfo {author} {\bibfnamefont {J.~A.}\ \bibnamefont
  {Ibers}},\ }\href@noop {} {\bibfield  {journal} {\bibinfo  {journal}
  {Inorganic Chemistry}\ }\textbf {\bibinfo {volume} {24}},\ \bibinfo {pages}
  {3611} (\bibinfo {year} {1985})}\BibitemShut {NoStop}%
\bibitem [{\citenamefont {Wakisaka}\ \emph {et~al.}(2009)\citenamefont
  {Wakisaka}, \citenamefont {Sudayama}, \citenamefont {Takubo}, \citenamefont
  {Mizokawa}, \citenamefont {Arita}, \citenamefont {Namatame}, \citenamefont
  {Taniguchi}, \citenamefont {Katayama}, \citenamefont {Nohara},\ and\
  \citenamefont {Takagi}}]{wakisaka2009excitonic}%
  \BibitemOpen
  \bibfield  {author} {\bibinfo {author} {\bibfnamefont {Y.}~\bibnamefont
  {Wakisaka}}, \bibinfo {author} {\bibfnamefont {T.}~\bibnamefont {Sudayama}},
  \bibinfo {author} {\bibfnamefont {K.}~\bibnamefont {Takubo}}, \bibinfo
  {author} {\bibfnamefont {T.}~\bibnamefont {Mizokawa}}, \bibinfo {author}
  {\bibfnamefont {M.}~\bibnamefont {Arita}}, \bibinfo {author} {\bibfnamefont
  {H.}~\bibnamefont {Namatame}}, \bibinfo {author} {\bibfnamefont
  {M.}~\bibnamefont {Taniguchi}}, \bibinfo {author} {\bibfnamefont
  {N.}~\bibnamefont {Katayama}}, \bibinfo {author} {\bibfnamefont
  {M.}~\bibnamefont {Nohara}}, \ and\ \bibinfo {author} {\bibfnamefont
  {H.}~\bibnamefont {Takagi}},\ }\href@noop {} {\bibfield  {journal} {\bibinfo
  {journal} {Physical review letters}\ }\textbf {\bibinfo {volume} {103}},\
  \bibinfo {pages} {026402} (\bibinfo {year} {2009})}\BibitemShut {NoStop}%
\bibitem [{\citenamefont {He}\ \emph {et~al.}(2021)\citenamefont {He},
  \citenamefont {Que}, \citenamefont {Zhou}, \citenamefont {Isobe},
  \citenamefont {Huang},\ and\ \citenamefont {Takagi}}]{he2021tunneling}%
  \BibitemOpen
  \bibfield  {author} {\bibinfo {author} {\bibfnamefont {Q.}~\bibnamefont
  {He}}, \bibinfo {author} {\bibfnamefont {X.}~\bibnamefont {Que}}, \bibinfo
  {author} {\bibfnamefont {L.}~\bibnamefont {Zhou}}, \bibinfo {author}
  {\bibfnamefont {M.}~\bibnamefont {Isobe}}, \bibinfo {author} {\bibfnamefont
  {D.}~\bibnamefont {Huang}}, \ and\ \bibinfo {author} {\bibfnamefont
  {H.}~\bibnamefont {Takagi}},\ }\href@noop {} {\bibfield  {journal} {\bibinfo
  {journal} {Physical Review Research}\ }\textbf {\bibinfo {volume} {3}},\
  \bibinfo {pages} {L032074} (\bibinfo {year} {2021})}\BibitemShut {NoStop}%
\bibitem [{\citenamefont {{Jia}}\ \emph {et~al.}(2020)\citenamefont {{Jia}},
  \citenamefont {{Wang}}, \citenamefont {{Chiu}}, \citenamefont {{Song}},
  \citenamefont {{Yu}}, \citenamefont {{J{\"a}ck}}, \citenamefont {{Lei}},
  \citenamefont {{Klemenz}}, \citenamefont {{Alexandre Cevallos}},
  \citenamefont {{Onyszczak}}, \citenamefont {{Fishchenko}}, \citenamefont
  {{Liu}}, \citenamefont {{Farahi}}, \citenamefont {{Xie}}, \citenamefont
  {{Xu}}, \citenamefont {{Watanabe}}, \citenamefont {{Taniguchi}},
  \citenamefont {{Bernevig}}, \citenamefont {{Cava}}, \citenamefont {{Schoop}},
  \citenamefont {{Yazdani}},\ and\ \citenamefont {{Wu}}}]{yanyu2020}%
  \BibitemOpen
  \bibfield  {author} {\bibinfo {author} {\bibfnamefont {Y.}~\bibnamefont
  {{Jia}}}, \bibinfo {author} {\bibfnamefont {P.}~\bibnamefont {{Wang}}},
  \bibinfo {author} {\bibfnamefont {C.-L.}\ \bibnamefont {{Chiu}}}, \bibinfo
  {author} {\bibfnamefont {Z.}~\bibnamefont {{Song}}}, \bibinfo {author}
  {\bibfnamefont {G.}~\bibnamefont {{Yu}}}, \bibinfo {author} {\bibfnamefont
  {B.}~\bibnamefont {{J{\"a}ck}}}, \bibinfo {author} {\bibfnamefont
  {S.}~\bibnamefont {{Lei}}}, \bibinfo {author} {\bibfnamefont
  {S.}~\bibnamefont {{Klemenz}}}, \bibinfo {author} {\bibfnamefont
  {F.}~\bibnamefont {{Alexandre Cevallos}}}, \bibinfo {author} {\bibfnamefont
  {M.}~\bibnamefont {{Onyszczak}}}, \bibinfo {author} {\bibfnamefont
  {N.}~\bibnamefont {{Fishchenko}}}, \bibinfo {author} {\bibfnamefont
  {X.}~\bibnamefont {{Liu}}}, \bibinfo {author} {\bibfnamefont
  {G.}~\bibnamefont {{Farahi}}}, \bibinfo {author} {\bibfnamefont
  {F.}~\bibnamefont {{Xie}}}, \bibinfo {author} {\bibfnamefont
  {Y.}~\bibnamefont {{Xu}}}, \bibinfo {author} {\bibfnamefont {K.}~\bibnamefont
  {{Watanabe}}}, \bibinfo {author} {\bibfnamefont {T.}~\bibnamefont
  {{Taniguchi}}}, \bibinfo {author} {\bibfnamefont {B.~A.}\ \bibnamefont
  {{Bernevig}}}, \bibinfo {author} {\bibfnamefont {R.~J.}\ \bibnamefont
  {{Cava}}}, \bibinfo {author} {\bibfnamefont {L.~M.}\ \bibnamefont
  {{Schoop}}}, \bibinfo {author} {\bibfnamefont {A.}~\bibnamefont {{Yazdani}}},
  \ and\ \bibinfo {author} {\bibfnamefont {S.}~\bibnamefont {{Wu}}},\
  }\href@noop {} {\bibfield  {journal} {\bibinfo  {journal} {arXiv e-prints}\
  ,\ \bibinfo {eid} {arXiv:2010.05390}} (\bibinfo {year} {2020})},\ \Eprint
  {http://arxiv.org/abs/2010.05390} {arXiv:2010.05390 [cond-mat.mes-hall]}
  \BibitemShut {NoStop}%
\bibitem [{\citenamefont {Li}\ \emph {et~al.}(2019)\citenamefont {Li},
  \citenamefont {Nadeem}, \citenamefont {Yue}, \citenamefont {Cortie},
  \citenamefont {Fuhrer},\ and\ \citenamefont {Wang}}]{li2019possible}%
  \BibitemOpen
  \bibfield  {author} {\bibinfo {author} {\bibfnamefont {Z.}~\bibnamefont
  {Li}}, \bibinfo {author} {\bibfnamefont {M.}~\bibnamefont {Nadeem}}, \bibinfo
  {author} {\bibfnamefont {Z.}~\bibnamefont {Yue}}, \bibinfo {author}
  {\bibfnamefont {D.}~\bibnamefont {Cortie}}, \bibinfo {author} {\bibfnamefont
  {M.}~\bibnamefont {Fuhrer}}, \ and\ \bibinfo {author} {\bibfnamefont
  {X.}~\bibnamefont {Wang}},\ }\href@noop {} {\bibfield  {journal} {\bibinfo
  {journal} {Nano letters}\ }\textbf {\bibinfo {volume} {19}},\ \bibinfo
  {pages} {4960} (\bibinfo {year} {2019})}\BibitemShut {NoStop}%
\bibitem [{\citenamefont {Wang}\ \emph {et~al.}(2019)\citenamefont {Wang},
  \citenamefont {Rhodes}, \citenamefont {Watanabe}, \citenamefont {Taniguchi},
  \citenamefont {Hone}, \citenamefont {Shan},\ and\ \citenamefont
  {Mak}}]{wang2019evidence}%
  \BibitemOpen
  \bibfield  {author} {\bibinfo {author} {\bibfnamefont {Z.}~\bibnamefont
  {Wang}}, \bibinfo {author} {\bibfnamefont {D.~A.}\ \bibnamefont {Rhodes}},
  \bibinfo {author} {\bibfnamefont {K.}~\bibnamefont {Watanabe}}, \bibinfo
  {author} {\bibfnamefont {T.}~\bibnamefont {Taniguchi}}, \bibinfo {author}
  {\bibfnamefont {J.~C.}\ \bibnamefont {Hone}}, \bibinfo {author}
  {\bibfnamefont {J.}~\bibnamefont {Shan}}, \ and\ \bibinfo {author}
  {\bibfnamefont {K.~F.}\ \bibnamefont {Mak}},\ }\href@noop {} {\bibfield
  {journal} {\bibinfo  {journal} {Nature}\ }\textbf {\bibinfo {volume} {574}},\
  \bibinfo {pages} {76} (\bibinfo {year} {2019})}\BibitemShut {NoStop}%
\bibitem [{\citenamefont {Watson}\ \emph {et~al.}(2020)\citenamefont {Watson},
  \citenamefont {Markovi\ifmmode~\acute{c}\else \'{c}\fi{}}, \citenamefont
  {Morales}, \citenamefont {Le~F\`evre}, \citenamefont {Merz}, \citenamefont
  {Haghighirad},\ and\ \citenamefont {King}}]{watson2020}%
  \BibitemOpen
  \bibfield  {author} {\bibinfo {author} {\bibfnamefont {M.~D.}\ \bibnamefont
  {Watson}}, \bibinfo {author} {\bibfnamefont {I.}~\bibnamefont
  {Markovi\ifmmode~\acute{c}\else \'{c}\fi{}}}, \bibinfo {author}
  {\bibfnamefont {E.~A.}\ \bibnamefont {Morales}}, \bibinfo {author}
  {\bibfnamefont {P.}~\bibnamefont {Le~F\`evre}}, \bibinfo {author}
  {\bibfnamefont {M.}~\bibnamefont {Merz}}, \bibinfo {author} {\bibfnamefont
  {A.~A.}\ \bibnamefont {Haghighirad}}, \ and\ \bibinfo {author} {\bibfnamefont
  {P.~D.~C.}\ \bibnamefont {King}},\ }\href {\doibase
  10.1103/PhysRevResearch.2.013236} {\bibfield  {journal} {\bibinfo  {journal}
  {Phys. Rev. Research}\ }\textbf {\bibinfo {volume} {2}},\ \bibinfo {pages}
  {013236} (\bibinfo {year} {2020})}\BibitemShut {NoStop}%
\bibitem [{\citenamefont {Werdehausen}\ \emph {et~al.}(2018)\citenamefont
  {Werdehausen}, \citenamefont {Takayama}, \citenamefont {H{\"o}ppner},
  \citenamefont {Albrecht}, \citenamefont {Rost}, \citenamefont {Lu},
  \citenamefont {Manske}, \citenamefont {Takagi},\ and\ \citenamefont
  {Kaiser}}]{werdehausen2018}%
  \BibitemOpen
  \bibfield  {author} {\bibinfo {author} {\bibfnamefont {D.}~\bibnamefont
  {Werdehausen}}, \bibinfo {author} {\bibfnamefont {T.}~\bibnamefont
  {Takayama}}, \bibinfo {author} {\bibfnamefont {M.}~\bibnamefont
  {H{\"o}ppner}}, \bibinfo {author} {\bibfnamefont {G.}~\bibnamefont
  {Albrecht}}, \bibinfo {author} {\bibfnamefont {A.~W.}\ \bibnamefont {Rost}},
  \bibinfo {author} {\bibfnamefont {Y.}~\bibnamefont {Lu}}, \bibinfo {author}
  {\bibfnamefont {D.}~\bibnamefont {Manske}}, \bibinfo {author} {\bibfnamefont
  {H.}~\bibnamefont {Takagi}}, \ and\ \bibinfo {author} {\bibfnamefont
  {S.}~\bibnamefont {Kaiser}},\ }\href@noop {} {\bibfield  {journal} {\bibinfo
  {journal} {Science advances}\ }\textbf {\bibinfo {volume} {4}},\ \bibinfo
  {pages} {eaap8652} (\bibinfo {year} {2018})}\BibitemShut {NoStop}%
\bibitem [{\citenamefont {Zenker}\ \emph {et~al.}(2014)\citenamefont {Zenker},
  \citenamefont {Fehske},\ and\ \citenamefont {Beck}}]{zenker2014}%
  \BibitemOpen
  \bibfield  {author} {\bibinfo {author} {\bibfnamefont {B.}~\bibnamefont
  {Zenker}}, \bibinfo {author} {\bibfnamefont {H.}~\bibnamefont {Fehske}}, \
  and\ \bibinfo {author} {\bibfnamefont {H.}~\bibnamefont {Beck}},\ }\href
  {\doibase 10.1103/PhysRevB.90.195118} {\bibfield  {journal} {\bibinfo
  {journal} {Phys. Rev. B}\ }\textbf {\bibinfo {volume} {90}},\ \bibinfo
  {pages} {195118} (\bibinfo {year} {2014})}\BibitemShut {NoStop}%
\bibitem [{\citenamefont {Kaneko}\ \emph {et~al.}(2012)\citenamefont {Kaneko},
  \citenamefont {Seki},\ and\ \citenamefont {Ohta}}]{kaneko2012}%
  \BibitemOpen
  \bibfield  {author} {\bibinfo {author} {\bibfnamefont {T.}~\bibnamefont
  {Kaneko}}, \bibinfo {author} {\bibfnamefont {K.}~\bibnamefont {Seki}}, \ and\
  \bibinfo {author} {\bibfnamefont {Y.}~\bibnamefont {Ohta}},\ }\href {\doibase
  10.1103/PhysRevB.85.165135} {\bibfield  {journal} {\bibinfo  {journal} {Phys.
  Rev. B}\ }\textbf {\bibinfo {volume} {85}},\ \bibinfo {pages} {165135}
  (\bibinfo {year} {2012})}\BibitemShut {NoStop}%
\bibitem [{\citenamefont {Seki}\ \emph {et~al.}(2014)\citenamefont {Seki},
  \citenamefont {Wakisaka}, \citenamefont {Kaneko}, \citenamefont {Toriyama},
  \citenamefont {Konishi}, \citenamefont {Sudayama}, \citenamefont {Saini},
  \citenamefont {Arita}, \citenamefont {Namatame}, \citenamefont {Taniguchi}
  \emph {et~al.}}]{seki2014excitonic}%
  \BibitemOpen
  \bibfield  {author} {\bibinfo {author} {\bibfnamefont {K.}~\bibnamefont
  {Seki}}, \bibinfo {author} {\bibfnamefont {Y.}~\bibnamefont {Wakisaka}},
  \bibinfo {author} {\bibfnamefont {T.}~\bibnamefont {Kaneko}}, \bibinfo
  {author} {\bibfnamefont {T.}~\bibnamefont {Toriyama}}, \bibinfo {author}
  {\bibfnamefont {T.}~\bibnamefont {Konishi}}, \bibinfo {author} {\bibfnamefont
  {T.}~\bibnamefont {Sudayama}}, \bibinfo {author} {\bibfnamefont
  {N.}~\bibnamefont {Saini}}, \bibinfo {author} {\bibfnamefont
  {M.}~\bibnamefont {Arita}}, \bibinfo {author} {\bibfnamefont
  {H.}~\bibnamefont {Namatame}}, \bibinfo {author} {\bibfnamefont
  {M.}~\bibnamefont {Taniguchi}},  \emph {et~al.},\ }\href@noop {} {\bibfield
  {journal} {\bibinfo  {journal} {Physical Review B}\ }\textbf {\bibinfo
  {volume} {90}},\ \bibinfo {pages} {155116} (\bibinfo {year}
  {2014})}\BibitemShut {NoStop}%
\bibitem [{\citenamefont {Butov}\ \emph {et~al.}(2002)\citenamefont {Butov},
  \citenamefont {Gossard},\ and\ \citenamefont {Chemla}}]{butov2002}%
  \BibitemOpen
  \bibfield  {author} {\bibinfo {author} {\bibfnamefont {L.}~\bibnamefont
  {Butov}}, \bibinfo {author} {\bibfnamefont {A.}~\bibnamefont {Gossard}}, \
  and\ \bibinfo {author} {\bibfnamefont {D.}~\bibnamefont {Chemla}},\
  }\href@noop {} {\bibfield  {journal} {\bibinfo  {journal} {Nature}\ }\textbf
  {\bibinfo {volume} {418}},\ \bibinfo {pages} {751} (\bibinfo {year}
  {2002})}\BibitemShut {NoStop}%
\bibitem [{\citenamefont {Sun}\ and\ \citenamefont
  {Millis}(2020)}]{sun2020BaSh}%
  \BibitemOpen
  \bibfield  {author} {\bibinfo {author} {\bibfnamefont {Z.}~\bibnamefont
  {Sun}}\ and\ \bibinfo {author} {\bibfnamefont {A.~J.}\ \bibnamefont
  {Millis}},\ }\href {\doibase 10.1103/PhysRevB.102.041110} {\bibfield
  {journal} {\bibinfo  {journal} {Phys. Rev. B}\ }\textbf {\bibinfo {volume}
  {102}},\ \bibinfo {pages} {041110} (\bibinfo {year} {2020})}\BibitemShut
  {NoStop}%
\bibitem [{\citenamefont {Nandi}\ \emph {et~al.}(2012)\citenamefont {Nandi},
  \citenamefont {Finck}, \citenamefont {Eisenstein}, \citenamefont {Pfeiffer},\
  and\ \citenamefont {West}}]{nandi2012exciton}%
  \BibitemOpen
  \bibfield  {author} {\bibinfo {author} {\bibfnamefont {D.}~\bibnamefont
  {Nandi}}, \bibinfo {author} {\bibfnamefont {A.}~\bibnamefont {Finck}},
  \bibinfo {author} {\bibfnamefont {J.}~\bibnamefont {Eisenstein}}, \bibinfo
  {author} {\bibfnamefont {L.}~\bibnamefont {Pfeiffer}}, \ and\ \bibinfo
  {author} {\bibfnamefont {K.}~\bibnamefont {West}},\ }\href@noop {} {\bibfield
   {journal} {\bibinfo  {journal} {Nature}\ }\textbf {\bibinfo {volume}
  {488}},\ \bibinfo {pages} {481} (\bibinfo {year} {2012})}\BibitemShut
  {NoStop}%
\bibitem [{\citenamefont {Ma}\ \emph {et~al.}(2022)\citenamefont {Ma},
  \citenamefont {Nguyen}, \citenamefont {Wang}, \citenamefont {Zeng},
  \citenamefont {Watanabe}, \citenamefont {Taniguchi}, \citenamefont
  {MacDonald}, \citenamefont {Mak},\ and\ \citenamefont
  {Shan}}]{ma2021strongly}%
  \BibitemOpen
  \bibfield  {author} {\bibinfo {author} {\bibfnamefont {L.}~\bibnamefont
  {Ma}}, \bibinfo {author} {\bibfnamefont {P.~X.}\ \bibnamefont {Nguyen}},
  \bibinfo {author} {\bibfnamefont {Z.}~\bibnamefont {Wang}}, \bibinfo {author}
  {\bibfnamefont {Y.}~\bibnamefont {Zeng}}, \bibinfo {author} {\bibfnamefont
  {K.}~\bibnamefont {Watanabe}}, \bibinfo {author} {\bibfnamefont
  {T.}~\bibnamefont {Taniguchi}}, \bibinfo {author} {\bibfnamefont {A.~H.}\
  \bibnamefont {MacDonald}}, \bibinfo {author} {\bibfnamefont {K.~F.}\
  \bibnamefont {Mak}}, \ and\ \bibinfo {author} {\bibfnamefont
  {J.}~\bibnamefont {Shan}},\ }\href@noop {} {\bibfield  {journal} {\bibinfo
  {journal} {Nature}\ }\textbf {\bibinfo {volume} {598}},\ \bibinfo {pages}
  {585–589} (\bibinfo {year} {2022})}\BibitemShut {NoStop}%
\bibitem [{\citenamefont {Seki}\ \emph {et~al.}(2011)\citenamefont {Seki},
  \citenamefont {Eder},\ and\ \citenamefont {Ohta}}]{seki2011}%
  \BibitemOpen
  \bibfield  {author} {\bibinfo {author} {\bibfnamefont {K.}~\bibnamefont
  {Seki}}, \bibinfo {author} {\bibfnamefont {R.}~\bibnamefont {Eder}}, \ and\
  \bibinfo {author} {\bibfnamefont {Y.}~\bibnamefont {Ohta}},\ }\href {\doibase
  10.1103/PhysRevB.84.245106} {\bibfield  {journal} {\bibinfo  {journal} {Phys.
  Rev. B}\ }\textbf {\bibinfo {volume} {84}},\ \bibinfo {pages} {245106}
  (\bibinfo {year} {2011})}\BibitemShut {NoStop}%
\bibitem [{\citenamefont {Mazza}\ \emph {et~al.}(2020)\citenamefont {Mazza},
  \citenamefont {R\"osner}, \citenamefont {Windg\"atter}, \citenamefont
  {Latini}, \citenamefont {H\"ubener}, \citenamefont {Millis}, \citenamefont
  {Rubio},\ and\ \citenamefont {Georges}}]{mazza2020}%
  \BibitemOpen
  \bibfield  {author} {\bibinfo {author} {\bibfnamefont {G.}~\bibnamefont
  {Mazza}}, \bibinfo {author} {\bibfnamefont {M.}~\bibnamefont {R\"osner}},
  \bibinfo {author} {\bibfnamefont {L.}~\bibnamefont {Windg\"atter}}, \bibinfo
  {author} {\bibfnamefont {S.}~\bibnamefont {Latini}}, \bibinfo {author}
  {\bibfnamefont {H.}~\bibnamefont {H\"ubener}}, \bibinfo {author}
  {\bibfnamefont {A.~J.}\ \bibnamefont {Millis}}, \bibinfo {author}
  {\bibfnamefont {A.}~\bibnamefont {Rubio}}, \ and\ \bibinfo {author}
  {\bibfnamefont {A.}~\bibnamefont {Georges}},\ }\href {\doibase
  10.1103/PhysRevLett.124.197601} {\bibfield  {journal} {\bibinfo  {journal}
  {Phys. Rev. Lett.}\ }\textbf {\bibinfo {volume} {124}},\ \bibinfo {pages}
  {197601} (\bibinfo {year} {2020})}\BibitemShut {NoStop}%
\bibitem [{\citenamefont {Kim}\ \emph {et~al.}(2021{\natexlab{b}})\citenamefont
  {Kim}, \citenamefont {Kim}, \citenamefont {Kim}, \citenamefont {Kwon},
  \citenamefont {Kim},\ and\ \citenamefont {Kim}}]{kim2021}%
  \BibitemOpen
  \bibfield  {author} {\bibinfo {author} {\bibfnamefont {K.}~\bibnamefont
  {Kim}}, \bibinfo {author} {\bibfnamefont {H.}~\bibnamefont {Kim}}, \bibinfo
  {author} {\bibfnamefont {J.}~\bibnamefont {Kim}}, \bibinfo {author}
  {\bibfnamefont {C.}~\bibnamefont {Kwon}}, \bibinfo {author} {\bibfnamefont
  {J.~S.}\ \bibnamefont {Kim}}, \ and\ \bibinfo {author} {\bibfnamefont
  {B.}~\bibnamefont {Kim}},\ }\href@noop {} {\bibfield  {journal} {\bibinfo
  {journal} {Nature communications}\ }\textbf {\bibinfo {volume} {12}},\
  \bibinfo {pages} {1} (\bibinfo {year} {2021}{\natexlab{b}})}\BibitemShut
  {NoStop}%
\bibitem [{\citenamefont {Larkin}\ \emph {et~al.}(2017)\citenamefont {Larkin},
  \citenamefont {Yaresko}, \citenamefont {Pr\"opper}, \citenamefont {Kikoin},
  \citenamefont {Lu}, \citenamefont {Takayama}, \citenamefont {Mathis},
  \citenamefont {Rost}, \citenamefont {Takagi}, \citenamefont {Keimer},\ and\
  \citenamefont {Boris}}]{larkin2017}%
  \BibitemOpen
  \bibfield  {author} {\bibinfo {author} {\bibfnamefont {T.~I.}\ \bibnamefont
  {Larkin}}, \bibinfo {author} {\bibfnamefont {A.~N.}\ \bibnamefont {Yaresko}},
  \bibinfo {author} {\bibfnamefont {D.}~\bibnamefont {Pr\"opper}}, \bibinfo
  {author} {\bibfnamefont {K.~A.}\ \bibnamefont {Kikoin}}, \bibinfo {author}
  {\bibfnamefont {Y.~F.}\ \bibnamefont {Lu}}, \bibinfo {author} {\bibfnamefont
  {T.}~\bibnamefont {Takayama}}, \bibinfo {author} {\bibfnamefont {Y.-L.}\
  \bibnamefont {Mathis}}, \bibinfo {author} {\bibfnamefont {A.~W.}\
  \bibnamefont {Rost}}, \bibinfo {author} {\bibfnamefont {H.}~\bibnamefont
  {Takagi}}, \bibinfo {author} {\bibfnamefont {B.}~\bibnamefont {Keimer}}, \
  and\ \bibinfo {author} {\bibfnamefont {A.~V.}\ \bibnamefont {Boris}},\ }\href
  {\doibase 10.1103/PhysRevB.95.195144} {\bibfield  {journal} {\bibinfo
  {journal} {Phys. Rev. B}\ }\textbf {\bibinfo {volume} {95}},\ \bibinfo
  {pages} {195144} (\bibinfo {year} {2017})}\BibitemShut {NoStop}%
\bibitem [{\citenamefont {Larkin}\ \emph {et~al.}(2018)\citenamefont {Larkin},
  \citenamefont {Dawson}, \citenamefont {H{\"o}ppner}, \citenamefont
  {Takayama}, \citenamefont {Isobe}, \citenamefont {Mathis}, \citenamefont
  {Takagi}, \citenamefont {Keimer},\ and\ \citenamefont {Boris}}]{larkin2018}%
  \BibitemOpen
  \bibfield  {author} {\bibinfo {author} {\bibfnamefont {T.}~\bibnamefont
  {Larkin}}, \bibinfo {author} {\bibfnamefont {R.}~\bibnamefont {Dawson}},
  \bibinfo {author} {\bibfnamefont {M.}~\bibnamefont {H{\"o}ppner}}, \bibinfo
  {author} {\bibfnamefont {T.}~\bibnamefont {Takayama}}, \bibinfo {author}
  {\bibfnamefont {M.}~\bibnamefont {Isobe}}, \bibinfo {author} {\bibfnamefont
  {Y.-L.}\ \bibnamefont {Mathis}}, \bibinfo {author} {\bibfnamefont
  {H.}~\bibnamefont {Takagi}}, \bibinfo {author} {\bibfnamefont
  {B.}~\bibnamefont {Keimer}}, \ and\ \bibinfo {author} {\bibfnamefont
  {A.}~\bibnamefont {Boris}},\ }\href@noop {} {\bibfield  {journal} {\bibinfo
  {journal} {Physical Review B}\ }\textbf {\bibinfo {volume} {98}},\ \bibinfo
  {pages} {125113} (\bibinfo {year} {2018})}\BibitemShut {NoStop}%
\bibitem [{\citenamefont {Bretscher}\ \emph {et~al.}(2021)\citenamefont
  {Bretscher}, \citenamefont {Andrich}, \citenamefont {Murakami}, \citenamefont
  {Gole{\v{z}}}, \citenamefont {Remez}, \citenamefont {Telang}, \citenamefont
  {Singh}, \citenamefont {Harnagea}, \citenamefont {Cooper}, \citenamefont
  {Millis} \emph {et~al.}}]{andrich2020}%
  \BibitemOpen
  \bibfield  {author} {\bibinfo {author} {\bibfnamefont {H.~M.}\ \bibnamefont
  {Bretscher}}, \bibinfo {author} {\bibfnamefont {P.}~\bibnamefont {Andrich}},
  \bibinfo {author} {\bibfnamefont {Y.}~\bibnamefont {Murakami}}, \bibinfo
  {author} {\bibfnamefont {D.}~\bibnamefont {Gole{\v{z}}}}, \bibinfo {author}
  {\bibfnamefont {B.}~\bibnamefont {Remez}}, \bibinfo {author} {\bibfnamefont
  {P.}~\bibnamefont {Telang}}, \bibinfo {author} {\bibfnamefont
  {A.}~\bibnamefont {Singh}}, \bibinfo {author} {\bibfnamefont
  {L.}~\bibnamefont {Harnagea}}, \bibinfo {author} {\bibfnamefont {N.~R.}\
  \bibnamefont {Cooper}}, \bibinfo {author} {\bibfnamefont {A.~J.}\
  \bibnamefont {Millis}},  \emph {et~al.},\ }\href@noop {} {\bibfield
  {journal} {\bibinfo  {journal} {Science Advances}\ }\textbf {\bibinfo
  {volume} {7}},\ \bibinfo {pages} {eabd6147} (\bibinfo {year}
  {2021})}\BibitemShut {NoStop}%
\bibitem [{\citenamefont {Okazaki}\ \emph {et~al.}(2018)\citenamefont
  {Okazaki}, \citenamefont {Ogawa}, \citenamefont {Suzuki}, \citenamefont
  {Yamamoto}, \citenamefont {Someya}, \citenamefont {Michimae}, \citenamefont
  {Watanabe}, \citenamefont {Lu}, \citenamefont {Nohara}, \citenamefont
  {Takagi} \emph {et~al.}}]{okazaki2018}%
  \BibitemOpen
  \bibfield  {author} {\bibinfo {author} {\bibfnamefont {K.}~\bibnamefont
  {Okazaki}}, \bibinfo {author} {\bibfnamefont {Y.}~\bibnamefont {Ogawa}},
  \bibinfo {author} {\bibfnamefont {T.}~\bibnamefont {Suzuki}}, \bibinfo
  {author} {\bibfnamefont {T.}~\bibnamefont {Yamamoto}}, \bibinfo {author}
  {\bibfnamefont {T.}~\bibnamefont {Someya}}, \bibinfo {author} {\bibfnamefont
  {S.}~\bibnamefont {Michimae}}, \bibinfo {author} {\bibfnamefont
  {M.}~\bibnamefont {Watanabe}}, \bibinfo {author} {\bibfnamefont
  {Y.}~\bibnamefont {Lu}}, \bibinfo {author} {\bibfnamefont {M.}~\bibnamefont
  {Nohara}}, \bibinfo {author} {\bibfnamefont {H.}~\bibnamefont {Takagi}},
  \emph {et~al.},\ }\href@noop {} {\bibfield  {journal} {\bibinfo  {journal}
  {Nature communications}\ }\textbf {\bibinfo {volume} {9}},\ \bibinfo {pages}
  {4322} (\bibinfo {year} {2018})}\BibitemShut {NoStop}%
\bibitem [{\citenamefont {Tang}\ \emph
  {et~al.}(2020{\natexlab{a}})\citenamefont {Tang}, \citenamefont {Wang},
  \citenamefont {Duan}, \citenamefont {Yang}, \citenamefont {Huang},
  \citenamefont {Guo}, \citenamefont {Qian},\ and\ \citenamefont
  {Zhang}}]{tang2020}%
  \BibitemOpen
  \bibfield  {author} {\bibinfo {author} {\bibfnamefont {T.}~\bibnamefont
  {Tang}}, \bibinfo {author} {\bibfnamefont {H.}~\bibnamefont {Wang}}, \bibinfo
  {author} {\bibfnamefont {S.}~\bibnamefont {Duan}}, \bibinfo {author}
  {\bibfnamefont {Y.}~\bibnamefont {Yang}}, \bibinfo {author} {\bibfnamefont
  {C.}~\bibnamefont {Huang}}, \bibinfo {author} {\bibfnamefont
  {Y.}~\bibnamefont {Guo}}, \bibinfo {author} {\bibfnamefont {D.}~\bibnamefont
  {Qian}}, \ and\ \bibinfo {author} {\bibfnamefont {W.}~\bibnamefont {Zhang}},\
  }\href {\doibase 10.1103/PhysRevB.101.235148} {\bibfield  {journal} {\bibinfo
   {journal} {Phys. Rev. B}\ }\textbf {\bibinfo {volume} {101}},\ \bibinfo
  {pages} {235148} (\bibinfo {year} {2020}{\natexlab{a}})}\BibitemShut
  {NoStop}%
\bibitem [{\citenamefont {Sugimoto}\ \emph {et~al.}(2018)\citenamefont
  {Sugimoto}, \citenamefont {Nishimoto}, \citenamefont {Kaneko},\ and\
  \citenamefont {Ohta}}]{sugimoto2018strong}%
  \BibitemOpen
  \bibfield  {author} {\bibinfo {author} {\bibfnamefont {K.}~\bibnamefont
  {Sugimoto}}, \bibinfo {author} {\bibfnamefont {S.}~\bibnamefont {Nishimoto}},
  \bibinfo {author} {\bibfnamefont {T.}~\bibnamefont {Kaneko}}, \ and\ \bibinfo
  {author} {\bibfnamefont {Y.}~\bibnamefont {Ohta}},\ }\href@noop {} {\bibfield
   {journal} {\bibinfo  {journal} {Physical review letters}\ }\textbf {\bibinfo
  {volume} {120}},\ \bibinfo {pages} {247602} (\bibinfo {year}
  {2018})}\BibitemShut {NoStop}%
\bibitem [{\citenamefont {Subedi}(2020)}]{subedi2020}%
  \BibitemOpen
  \bibfield  {author} {\bibinfo {author} {\bibfnamefont {A.}~\bibnamefont
  {Subedi}},\ }\href {\doibase 10.1103/PhysRevMaterials.4.083601} {\bibfield
  {journal} {\bibinfo  {journal} {Phys. Rev. Materials}\ }\textbf {\bibinfo
  {volume} {4}},\ \bibinfo {pages} {083601} (\bibinfo {year}
  {2020})}\BibitemShut {NoStop}%
\bibitem [{\citenamefont {Windgätter}\ \emph {et~al.}(2021)\citenamefont
  {Windgätter}, \citenamefont {Rösner}, \citenamefont {Mazza}, \citenamefont
  {Hübener}, \citenamefont {Georges}, \citenamefont {Millis}, \citenamefont
  {Latini},\ and\ \citenamefont {Rubio}}]{windgatter2021}%
  \BibitemOpen
  \bibfield  {author} {\bibinfo {author} {\bibfnamefont {L.}~\bibnamefont
  {Windgätter}}, \bibinfo {author} {\bibfnamefont {M.}~\bibnamefont
  {Rösner}}, \bibinfo {author} {\bibfnamefont {G.}~\bibnamefont {Mazza}},
  \bibinfo {author} {\bibfnamefont {H.}~\bibnamefont {Hübener}}, \bibinfo
  {author} {\bibfnamefont {A.}~\bibnamefont {Georges}}, \bibinfo {author}
  {\bibfnamefont {A.~J.}\ \bibnamefont {Millis}}, \bibinfo {author}
  {\bibfnamefont {S.}~\bibnamefont {Latini}}, \ and\ \bibinfo {author}
  {\bibfnamefont {A.}~\bibnamefont {Rubio}},\ }\href@noop {} {\enquote
  {\bibinfo {title} {Common microscopic origin of the phase transitions in
  ta$_2$nis$_5$ and the excitonic insulator candidate ta$_2$nise$_5$},}\ }
  (\bibinfo {year} {2021}),\ \Eprint {http://arxiv.org/abs/2105.13924}
  {arXiv:2105.13924 [cond-mat.mtrl-sci]} \BibitemShut {NoStop}%
\bibitem [{\citenamefont {Gole\ifmmode~\check{z}\else \v{z}\fi{}}\ \emph
  {et~al.}(2020)\citenamefont {Gole\ifmmode~\check{z}\else \v{z}\fi{}},
  \citenamefont {Sun}, \citenamefont {Murakami}, \citenamefont {Georges},\ and\
  \citenamefont {Millis}}]{golez2020}%
  \BibitemOpen
  \bibfield  {author} {\bibinfo {author} {\bibfnamefont {D.}~\bibnamefont
  {Gole\ifmmode~\check{z}\else \v{z}\fi{}}}, \bibinfo {author} {\bibfnamefont
  {Z.}~\bibnamefont {Sun}}, \bibinfo {author} {\bibfnamefont {Y.}~\bibnamefont
  {Murakami}}, \bibinfo {author} {\bibfnamefont {A.}~\bibnamefont {Georges}}, \
  and\ \bibinfo {author} {\bibfnamefont {A.~J.}\ \bibnamefont {Millis}},\
  }\href {\doibase 10.1103/PhysRevLett.125.257601} {\bibfield  {journal}
  {\bibinfo  {journal} {Phys. Rev. Lett.}\ }\textbf {\bibinfo {volume} {125}},\
  \bibinfo {pages} {257601} (\bibinfo {year} {2020})}\BibitemShut {NoStop}%
\bibitem [{\citenamefont {Murakami}\ \emph {et~al.}(2020)\citenamefont
  {Murakami}, \citenamefont {Gole\ifmmode~\check{z}\else \v{z}\fi{}},
  \citenamefont {Kaneko}, \citenamefont {Koga}, \citenamefont {Millis},\ and\
  \citenamefont {Werner}}]{murakami2020}%
  \BibitemOpen
  \bibfield  {author} {\bibinfo {author} {\bibfnamefont {Y.}~\bibnamefont
  {Murakami}}, \bibinfo {author} {\bibfnamefont {D.}~\bibnamefont
  {Gole\ifmmode~\check{z}\else \v{z}\fi{}}}, \bibinfo {author} {\bibfnamefont
  {T.}~\bibnamefont {Kaneko}}, \bibinfo {author} {\bibfnamefont
  {A.}~\bibnamefont {Koga}}, \bibinfo {author} {\bibfnamefont {A.~J.}\
  \bibnamefont {Millis}}, \ and\ \bibinfo {author} {\bibfnamefont
  {P.}~\bibnamefont {Werner}},\ }\href {\doibase 10.1103/PhysRevB.101.195118}
  {\bibfield  {journal} {\bibinfo  {journal} {Phys. Rev. B}\ }\textbf {\bibinfo
  {volume} {101}},\ \bibinfo {pages} {195118} (\bibinfo {year}
  {2020})}\BibitemShut {NoStop}%
\bibitem [{\citenamefont {Murakami}\ \emph {et~al.}(2017)\citenamefont
  {Murakami}, \citenamefont {Gole\ifmmode~\check{z}\else \v{z}\fi{}},
  \citenamefont {Eckstein},\ and\ \citenamefont {Werner}}]{murakami2017Photo}%
  \BibitemOpen
  \bibfield  {author} {\bibinfo {author} {\bibfnamefont {Y.}~\bibnamefont
  {Murakami}}, \bibinfo {author} {\bibfnamefont {D.}~\bibnamefont
  {Gole\ifmmode~\check{z}\else \v{z}\fi{}}}, \bibinfo {author} {\bibfnamefont
  {M.}~\bibnamefont {Eckstein}}, \ and\ \bibinfo {author} {\bibfnamefont
  {P.}~\bibnamefont {Werner}},\ }\href
  {https://link.aps.org/doi/10.1103/PhysRevLett.119.247601} {\bibfield
  {journal} {\bibinfo  {journal} {Phys. Rev. Lett.}\ }\textbf {\bibinfo
  {volume} {119}},\ \bibinfo {pages} {247601} (\bibinfo {year}
  {2017})}\BibitemShut {NoStop}%
\bibitem [{\citenamefont {Tanaka}\ \emph {et~al.}(2018)\citenamefont {Tanaka},
  \citenamefont {Daira},\ and\ \citenamefont {Yonemitsu}}]{tanaka2018}%
  \BibitemOpen
  \bibfield  {author} {\bibinfo {author} {\bibfnamefont {Y.}~\bibnamefont
  {Tanaka}}, \bibinfo {author} {\bibfnamefont {M.}~\bibnamefont {Daira}}, \
  and\ \bibinfo {author} {\bibfnamefont {K.}~\bibnamefont {Yonemitsu}},\
  }\href@noop {} {\bibfield  {journal} {\bibinfo  {journal} {Physical Review
  B}\ }\textbf {\bibinfo {volume} {97}},\ \bibinfo {pages} {115105} (\bibinfo
  {year} {2018})}\BibitemShut {NoStop}%
\bibitem [{\citenamefont {Tanabe}\ \emph {et~al.}(2018)\citenamefont {Tanabe},
  \citenamefont {Sugimoto},\ and\ \citenamefont {Ohta}}]{tanabe2018}%
  \BibitemOpen
  \bibfield  {author} {\bibinfo {author} {\bibfnamefont {T.}~\bibnamefont
  {Tanabe}}, \bibinfo {author} {\bibfnamefont {K.}~\bibnamefont {Sugimoto}}, \
  and\ \bibinfo {author} {\bibfnamefont {Y.}~\bibnamefont {Ohta}},\ }\href@noop
  {} {\bibfield  {journal} {\bibinfo  {journal} {Physical Review B}\ }\textbf
  {\bibinfo {volume} {98}},\ \bibinfo {pages} {235127} (\bibinfo {year}
  {2018})}\BibitemShut {NoStop}%
\bibitem [{\citenamefont {Fujiuchi}\ \emph {et~al.}(2019)\citenamefont
  {Fujiuchi}, \citenamefont {Kaneko}, \citenamefont {Ohta},\ and\ \citenamefont
  {Yunoki}}]{ryo2019}%
  \BibitemOpen
  \bibfield  {author} {\bibinfo {author} {\bibfnamefont {R.}~\bibnamefont
  {Fujiuchi}}, \bibinfo {author} {\bibfnamefont {T.}~\bibnamefont {Kaneko}},
  \bibinfo {author} {\bibfnamefont {Y.}~\bibnamefont {Ohta}}, \ and\ \bibinfo
  {author} {\bibfnamefont {S.}~\bibnamefont {Yunoki}},\ }\href {\doibase
  10.1103/PhysRevB.100.045121} {\bibfield  {journal} {\bibinfo  {journal}
  {Phys. Rev. B}\ }\textbf {\bibinfo {volume} {100}},\ \bibinfo {pages}
  {045121} (\bibinfo {year} {2019})}\BibitemShut {NoStop}%
\bibitem [{\citenamefont {Boschini}\ \emph {et~al.}(2018)\citenamefont
  {Boschini}, \citenamefont {da~Silva~Neto}, \citenamefont {Razzoli},
  \citenamefont {Zonno}, \citenamefont {Peli}, \citenamefont {Day},
  \citenamefont {Michiardi}, \citenamefont {Schneider}, \citenamefont
  {Zwartsenberg}, \citenamefont {Nigge}, \citenamefont {Zhong}, \citenamefont
  {Schneeloch}, \citenamefont {Gu}, \citenamefont {Zhdanovich}, \citenamefont
  {Mills}, \citenamefont {Levy}, \citenamefont {Jones}, \citenamefont
  {Giannetti},\ and\ \citenamefont {Damascelli}}]{Boshini2018}%
  \BibitemOpen
  \bibfield  {author} {\bibinfo {author} {\bibfnamefont {F.}~\bibnamefont
  {Boschini}}, \bibinfo {author} {\bibfnamefont {E.~H.}\ \bibnamefont
  {da~Silva~Neto}}, \bibinfo {author} {\bibfnamefont {E.}~\bibnamefont
  {Razzoli}}, \bibinfo {author} {\bibfnamefont {M.}~\bibnamefont {Zonno}},
  \bibinfo {author} {\bibfnamefont {S.}~\bibnamefont {Peli}}, \bibinfo {author}
  {\bibfnamefont {R.~P.}\ \bibnamefont {Day}}, \bibinfo {author} {\bibfnamefont
  {M.}~\bibnamefont {Michiardi}}, \bibinfo {author} {\bibfnamefont
  {M.}~\bibnamefont {Schneider}}, \bibinfo {author} {\bibfnamefont
  {B.}~\bibnamefont {Zwartsenberg}}, \bibinfo {author} {\bibfnamefont
  {P.}~\bibnamefont {Nigge}}, \bibinfo {author} {\bibfnamefont {R.~D.}\
  \bibnamefont {Zhong}}, \bibinfo {author} {\bibfnamefont {J.}~\bibnamefont
  {Schneeloch}}, \bibinfo {author} {\bibfnamefont {G.~D.}\ \bibnamefont {Gu}},
  \bibinfo {author} {\bibfnamefont {S.}~\bibnamefont {Zhdanovich}}, \bibinfo
  {author} {\bibfnamefont {A.~K.}\ \bibnamefont {Mills}}, \bibinfo {author}
  {\bibfnamefont {G.}~\bibnamefont {Levy}}, \bibinfo {author} {\bibfnamefont
  {D.~J.}\ \bibnamefont {Jones}}, \bibinfo {author} {\bibfnamefont
  {C.}~\bibnamefont {Giannetti}}, \ and\ \bibinfo {author} {\bibfnamefont
  {A.}~\bibnamefont {Damascelli}},\ }\href {\doibase 10.1038/s41563-018-0045-1}
  {\bibfield  {journal} {\bibinfo  {journal} {Nature Materials}\ }\textbf
  {\bibinfo {volume} {17}},\ \bibinfo {pages} {416} (\bibinfo {year}
  {2018})}\BibitemShut {NoStop}%
\bibitem [{\citenamefont {Kaneko}\ and\ \citenamefont
  {Ohta}(2014)}]{kaneko2014a}%
  \BibitemOpen
  \bibfield  {author} {\bibinfo {author} {\bibfnamefont {T.}~\bibnamefont
  {Kaneko}}\ and\ \bibinfo {author} {\bibfnamefont {Y.}~\bibnamefont {Ohta}},\
  }\href {\doibase 10.7566/JPSJ.83.024711} {\bibfield  {journal} {\bibinfo
  {journal} {Journal of the Physical Society of Japan}\ }\textbf {\bibinfo
  {volume} {83}},\ \bibinfo {pages} {024711} (\bibinfo {year} {2014})},\
  \Eprint {http://arxiv.org/abs/https://doi.org/10.7566/JPSJ.83.024711}
  {https://doi.org/10.7566/JPSJ.83.024711} \BibitemShut {NoStop}%
\bibitem [{\citenamefont {Zenker}\ \emph {et~al.}(2012)\citenamefont {Zenker},
  \citenamefont {Ihle}, \citenamefont {Bronold},\ and\ \citenamefont
  {Fehske}}]{zenker2012}%
  \BibitemOpen
  \bibfield  {author} {\bibinfo {author} {\bibfnamefont {B.}~\bibnamefont
  {Zenker}}, \bibinfo {author} {\bibfnamefont {D.}~\bibnamefont {Ihle}},
  \bibinfo {author} {\bibfnamefont {F.~X.}\ \bibnamefont {Bronold}}, \ and\
  \bibinfo {author} {\bibfnamefont {H.}~\bibnamefont {Fehske}},\ }\href
  {\doibase 10.1103/PhysRevB.85.121102} {\bibfield  {journal} {\bibinfo
  {journal} {Phys. Rev. B}\ }\textbf {\bibinfo {volume} {85}},\ \bibinfo
  {pages} {121102} (\bibinfo {year} {2012})}\BibitemShut {NoStop}%
\bibitem [{SM()}]{SM}%
  \BibitemOpen
  \href@noop {} {\enquote {\bibinfo {title} {Supplemental material includes (i)
  detailed description of the theoretical model, the evaluation of the
  photo-emission spectrum and method, (ii) analysis of photo-emission changes
  at higher momenta, (iii) experimental data analysis and photo-doping
  estimations. {A}dditional references included are
  [\onlinecite{li2020}],~[\onlinecite{freericks2009}],~[\onlinecite{eckstein2008}].}}\
  }\BibitemShut {NoStop}%
\bibitem [{\citenamefont {Li}\ \emph {et~al.}(2020)\citenamefont {Li},
  \citenamefont {Golez}, \citenamefont {Mazza}, \citenamefont {Millis},
  \citenamefont {Georges},\ and\ \citenamefont {Eckstein}}]{li2020}%
  \BibitemOpen
  \bibfield  {author} {\bibinfo {author} {\bibfnamefont {J.}~\bibnamefont
  {Li}}, \bibinfo {author} {\bibfnamefont {D.}~\bibnamefont {Golez}}, \bibinfo
  {author} {\bibfnamefont {G.}~\bibnamefont {Mazza}}, \bibinfo {author}
  {\bibfnamefont {A.~J.}\ \bibnamefont {Millis}}, \bibinfo {author}
  {\bibfnamefont {A.}~\bibnamefont {Georges}}, \ and\ \bibinfo {author}
  {\bibfnamefont {M.}~\bibnamefont {Eckstein}},\ }\href {\doibase
  10.1103/PhysRevB.101.205140} {\bibfield  {journal} {\bibinfo  {journal}
  {Phys. Rev. B}\ }\textbf {\bibinfo {volume} {101}},\ \bibinfo {pages}
  {205140} (\bibinfo {year} {2020})}\BibitemShut {NoStop}%
\bibitem [{\citenamefont {Gole\ifmmode~\check{z}\else \v{z}\fi{}}\ \emph
  {et~al.}(2019)\citenamefont {Gole\ifmmode~\check{z}\else \v{z}\fi{}},
  \citenamefont {Eckstein},\ and\ \citenamefont {Werner}}]{golez2019}%
  \BibitemOpen
  \bibfield  {author} {\bibinfo {author} {\bibfnamefont {D.}~\bibnamefont
  {Gole\ifmmode~\check{z}\else \v{z}\fi{}}}, \bibinfo {author} {\bibfnamefont
  {M.}~\bibnamefont {Eckstein}}, \ and\ \bibinfo {author} {\bibfnamefont
  {P.}~\bibnamefont {Werner}},\ }\href {\doibase 10.1103/PhysRevB.100.235117}
  {\bibfield  {journal} {\bibinfo  {journal} {Phys. Rev. B}\ }\textbf {\bibinfo
  {volume} {100}},\ \bibinfo {pages} {235117} (\bibinfo {year}
  {2019})}\BibitemShut {NoStop}%
\bibitem [{\citenamefont {Dmytruk}\ and\ \citenamefont
  {Schir\'o}(2021)}]{dmytruk2020gauge}%
  \BibitemOpen
  \bibfield  {author} {\bibinfo {author} {\bibfnamefont {O.}~\bibnamefont
  {Dmytruk}}\ and\ \bibinfo {author} {\bibfnamefont {M.}~\bibnamefont
  {Schir\'o}},\ }\href {\doibase 10.1103/PhysRevB.103.075131} {\bibfield
  {journal} {\bibinfo  {journal} {Phys. Rev. B}\ }\textbf {\bibinfo {volume}
  {103}},\ \bibinfo {pages} {075131} (\bibinfo {year} {2021})}\BibitemShut
  {NoStop}%
\bibitem [{\citenamefont {Saha}\ \emph {et~al.}(2021)\citenamefont {Saha},
  \citenamefont {Gole\ifmmode~\check{z}\else \v{z}\fi{}}, \citenamefont
  {De~Ninno}, \citenamefont {Mravlje}, \citenamefont {Murakami}, \citenamefont
  {Ressel}, \citenamefont {Stupar},\ and\ \citenamefont
  {Ribi\ifmmode~\check{c}\else \v{c}\fi{}}}]{saha2021}%
  \BibitemOpen
  \bibfield  {author} {\bibinfo {author} {\bibfnamefont {T.}~\bibnamefont
  {Saha}}, \bibinfo {author} {\bibfnamefont {D.}~\bibnamefont
  {Gole\ifmmode~\check{z}\else \v{z}\fi{}}}, \bibinfo {author} {\bibfnamefont
  {G.}~\bibnamefont {De~Ninno}}, \bibinfo {author} {\bibfnamefont
  {J.}~\bibnamefont {Mravlje}}, \bibinfo {author} {\bibfnamefont
  {Y.}~\bibnamefont {Murakami}}, \bibinfo {author} {\bibfnamefont
  {B.}~\bibnamefont {Ressel}}, \bibinfo {author} {\bibfnamefont
  {M.}~\bibnamefont {Stupar}}, \ and\ \bibinfo {author} {\bibfnamefont {P.~c.
  v.~R.}\ \bibnamefont {Ribi\ifmmode~\check{c}\else \v{c}\fi{}}},\ }\href
  {\doibase 10.1103/PhysRevB.103.144304} {\bibfield  {journal} {\bibinfo
  {journal} {Phys. Rev. B}\ }\textbf {\bibinfo {volume} {103}},\ \bibinfo
  {pages} {144304} (\bibinfo {year} {2021})}\BibitemShut {NoStop}%
\bibitem [{\citenamefont {Matsubayashi}\ \emph {et~al.}(2021)\citenamefont
  {Matsubayashi}, \citenamefont {Okamura}, \citenamefont {Mizokawa},
  \citenamefont {Katayama}, \citenamefont {Nakano}, \citenamefont {Sawa},
  \citenamefont {Kaneko}, \citenamefont {Toriyama}, \citenamefont {Konishi},
  \citenamefont {Ohta} \emph {et~al.}}]{matsubayashi2021hybridization}%
  \BibitemOpen
  \bibfield  {author} {\bibinfo {author} {\bibfnamefont {K.}~\bibnamefont
  {Matsubayashi}}, \bibinfo {author} {\bibfnamefont {H.}~\bibnamefont
  {Okamura}}, \bibinfo {author} {\bibfnamefont {T.}~\bibnamefont {Mizokawa}},
  \bibinfo {author} {\bibfnamefont {N.}~\bibnamefont {Katayama}}, \bibinfo
  {author} {\bibfnamefont {A.}~\bibnamefont {Nakano}}, \bibinfo {author}
  {\bibfnamefont {H.}~\bibnamefont {Sawa}}, \bibinfo {author} {\bibfnamefont
  {T.}~\bibnamefont {Kaneko}}, \bibinfo {author} {\bibfnamefont
  {T.}~\bibnamefont {Toriyama}}, \bibinfo {author} {\bibfnamefont
  {T.}~\bibnamefont {Konishi}}, \bibinfo {author} {\bibfnamefont
  {Y.}~\bibnamefont {Ohta}},  \emph {et~al.},\ }\href@noop {} {\bibfield
  {journal} {\bibinfo  {journal} {Journal of the Physical Society of Japan}\
  }\textbf {\bibinfo {volume} {90}},\ \bibinfo {pages} {074706} (\bibinfo
  {year} {2021})}\BibitemShut {NoStop}%
\bibitem [{\citenamefont {Sch{\"u}ler}\ \emph {et~al.}(2020)\citenamefont
  {Sch{\"u}ler}, \citenamefont {Gole{\v{z}}}, \citenamefont {Murakami},
  \citenamefont {Bittner}, \citenamefont {Herrmann}, \citenamefont {Strand},
  \citenamefont {Werner},\ and\ \citenamefont {Eckstein}}]{schuler2020nessi}%
  \BibitemOpen
  \bibfield  {author} {\bibinfo {author} {\bibfnamefont {M.}~\bibnamefont
  {Sch{\"u}ler}}, \bibinfo {author} {\bibfnamefont {D.}~\bibnamefont
  {Gole{\v{z}}}}, \bibinfo {author} {\bibfnamefont {Y.}~\bibnamefont
  {Murakami}}, \bibinfo {author} {\bibfnamefont {N.}~\bibnamefont {Bittner}},
  \bibinfo {author} {\bibfnamefont {A.}~\bibnamefont {Herrmann}}, \bibinfo
  {author} {\bibfnamefont {H.~U.}\ \bibnamefont {Strand}}, \bibinfo {author}
  {\bibfnamefont {P.}~\bibnamefont {Werner}}, \ and\ \bibinfo {author}
  {\bibfnamefont {M.}~\bibnamefont {Eckstein}},\ }\href@noop {} {\bibfield
  {journal} {\bibinfo  {journal} {Computer Physics Communications}\ ,\ \bibinfo
  {pages} {107484}} (\bibinfo {year} {2020})}\BibitemShut {NoStop}%
\bibitem [{\citenamefont {Freericks}\ \emph {et~al.}(2009)\citenamefont
  {Freericks}, \citenamefont {Krishnamurthy},\ and\ \citenamefont
  {Pruschke}}]{freericks2009}%
  \BibitemOpen
  \bibfield  {author} {\bibinfo {author} {\bibfnamefont {J.~K.}\ \bibnamefont
  {Freericks}}, \bibinfo {author} {\bibfnamefont {H.~R.}\ \bibnamefont
  {Krishnamurthy}}, \ and\ \bibinfo {author} {\bibfnamefont {T.}~\bibnamefont
  {Pruschke}},\ }\href {\doibase 10.1103/PhysRevLett.102.136401} {\bibfield
  {journal} {\bibinfo  {journal} {Phys. Rev. Lett.}\ }\textbf {\bibinfo
  {volume} {102}},\ \bibinfo {pages} {136401} (\bibinfo {year}
  {2009})}\BibitemShut {NoStop}%
\bibitem [{\citenamefont {Eckstein}\ and\ \citenamefont
  {Kollar}(2008)}]{eckstein2008}%
  \BibitemOpen
  \bibfield  {author} {\bibinfo {author} {\bibfnamefont {M.}~\bibnamefont
  {Eckstein}}\ and\ \bibinfo {author} {\bibfnamefont {M.}~\bibnamefont
  {Kollar}},\ }\href {\doibase 10.1103/PhysRevB.78.205119} {\bibfield
  {journal} {\bibinfo  {journal} {Phys. Rev. B}\ }\textbf {\bibinfo {volume}
  {78}},\ \bibinfo {pages} {205119} (\bibinfo {year} {2008})}\BibitemShut
  {NoStop}%
\bibitem [{\citenamefont {Tang}\ \emph
  {et~al.}(2020{\natexlab{b}})\citenamefont {Tang}, \citenamefont {Wang},
  \citenamefont {Duan}, \citenamefont {Yang}, \citenamefont {Huang},
  \citenamefont {Guo}, \citenamefont {Qian},\ and\ \citenamefont
  {Zhang}}]{bandshiftTang}%
  \BibitemOpen
  \bibfield  {author} {\bibinfo {author} {\bibfnamefont {T.}~\bibnamefont
  {Tang}}, \bibinfo {author} {\bibfnamefont {H.}~\bibnamefont {Wang}}, \bibinfo
  {author} {\bibfnamefont {S.}~\bibnamefont {Duan}}, \bibinfo {author}
  {\bibfnamefont {Y.}~\bibnamefont {Yang}}, \bibinfo {author} {\bibfnamefont
  {C.}~\bibnamefont {Huang}}, \bibinfo {author} {\bibfnamefont
  {Y.}~\bibnamefont {Guo}}, \bibinfo {author} {\bibfnamefont {D.}~\bibnamefont
  {Qian}}, \ and\ \bibinfo {author} {\bibfnamefont {W.}~\bibnamefont {Zhang}},\
  }\href {\doibase 10.1103/PhysRevB.101.235148} {\bibfield  {journal} {\bibinfo
   {journal} {Phys. Rev. B}\ }\textbf {\bibinfo {volume} {101}},\ \bibinfo
  {pages} {235148} (\bibinfo {year} {2020}{\natexlab{b}})}\BibitemShut
  {NoStop}%
\end{thebibliography}%

\appendix
\clearpage

\section{Model}\label{SM:Model}
\vspace{-0.05 cm}
In this section, we describe a mapping from the ab-initio determined bandstructure in Ref.~\onlinecite{mazza2020} to the minimal two-band  model employed in the main text. The total Hamiltonian consist of the kinetic and the interaction part as
\beq{
H=H_{\text{kin}}+H_{\text{int}}.
}
For the kinetic part, we will consider a  single Ta-Ni-Ta chain and neglect the interchain hopping, as their energy scales are much smaller than the energy scales of the gap size considered in this study. The tight-binding parameterization of the real-space Hamiltonian is given by~\cite{mazza2020}
\begin{equation*}
\begin{split}
    H_{\text{kin}}=&\sum_{\substack{i,\alpha\in\{1,2\}}}\left[  t_{T} c_{T,\alpha,i}^{\dagger} c_{T,\alpha,i+1} +\epsilon_{T} c_{T,\alpha,i}^{\dagger} c_{T,\alpha,i}+h.c.\right]+ \\
    &\sum_{\substack{i,\alpha\in\{1,2\}}} t_{TN}\left[c_{T,\alpha,i+1}^{\dagger}c_{N,i}-
    c_{T,\alpha,i}^{\dagger} c_{N,i}+h.c.\right]+    \\
    &\hspace{0.3cm}\sum_i  \left[t_{N} c_{N,i}^{\dagger} c_{N,i+1}+\epsilon_{N} c_{N,i}^{\dagger} c_{N,i} +h.c.\right]-\\
    &E(t) \sum_i c^{\dagger}_{T,\alpha,i} c_{N,i}+h.c.,\\
\end{split}
\end{equation*}
where $c_{T,\alpha,i},c_{N,i}$ are annihilation operators at site $i$ for Ta orbital in $\alpha\in\{1,2\}$ chain and Ni orbital, respectively.  The hoppings are given by $t_{T}=-0.72$ eV, $t_{N}=0.3$ eV and the crystal fields are $\epsilon_{N}=-0.36$ eV and $\epsilon_{T}=1.35$ eV.

The interaction part is determined by the competition between electronic and lattice degree of freedom as 
\bsplit{
H_{\text{int}}= V& \sum_{\substack{i,\alpha\in\{0,1\}}} n_{T,\alpha,i} n_{N,i}+ \sum_i \frac{1}{2} [ X^2_i+\frac{1}{\omega_0^2} \dot{{ X_i^2}}] \\
 &\sum_{\substack{i,\alpha\in\{0,1\}}}\sqrt{\lambda}  X_i \text{ + h.c.},
}
where $n_{T,\alpha,i}(n_{N,i})$ are density operators at site i for Ta~(Ni) orbitals and $\hat X$ is the distortion operator.

\begin{figure}[t]
\includegraphics[width=1.0\linewidth]{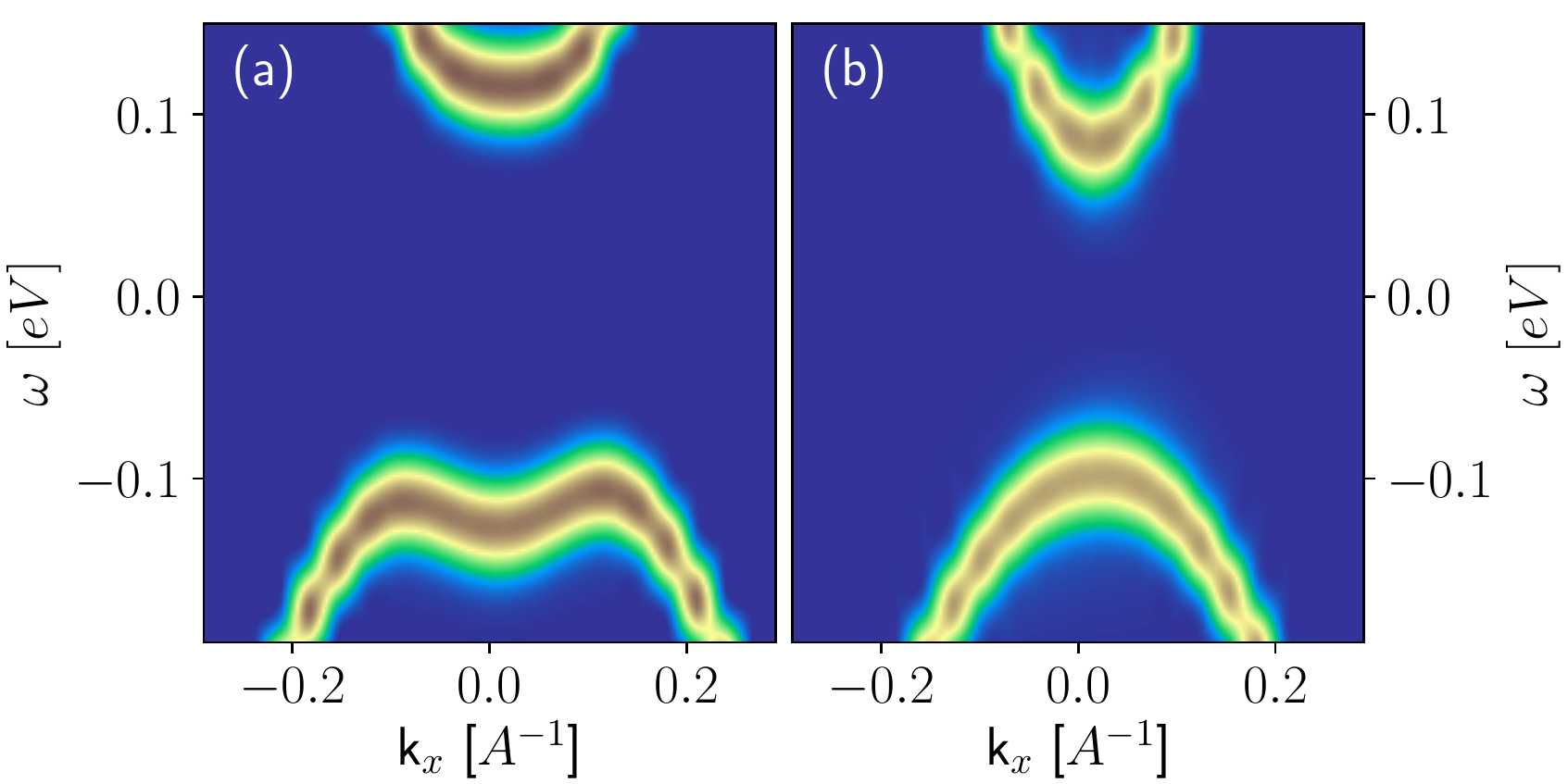}
\caption{Equilibrium momentum-dependent spectral function in the electron dominated regime for (a)~semimetallic~(BCS) regime~($V=0.76$ eV, $\lambda=0.03$ eV and $\delta=0.36$ eV) and (b)~semiconducting~(BEC) regime~($V=0.76$ eV, $\lambda=0.03$ eV and $\delta=0.51$ eV).}
\label{Fig:app1}
\end{figure}

The BCS-BEC transition is driven by the crystal-field splitting $\delta$ determined as $\epsilon_{N}\rightarrow\epsilon_{N}+\delta/2~(\epsilon_{T}\rightarrow\epsilon_{T}-\delta/2).$ In the BCS regime, we have adjusted the crystal-field splitting by  for $\delta=$0.36 eV to to fix the maximum in the "M"-shaped valence band to the experimentally relevant position, namely $k$=0.08 A$^{-1},$ see Fig.~\ref{Fig:exp1}(a). The BEC example presented in the main text corresponds to $\delta=$0.51 eV, see Fig.~\ref{Fig:app1}(b). The equilibrium spectrum in the lattice dominated regime is barely distinguishable from the electronic driven examples.

Performing the Fourier transform for Ta chains as $c_{T,1,k}(c_{T,2,k})=\sum_j e^{\I k j} c_{T,1,j}(c_{T,2,j})$ and with a shift for Ni chains $c_{N,k}=\sum_j e^{\I k (j-1/2)} c_{N,j}$, we obtain the tight-binding Hamiltonian in momentum space as
\begin{widetext}
\bsplit{
	&H=\sum_k  \psi_k^{\dagger} \begin{pmatrix}
	-2t_{T} \cos(k_x) +\epsilon_{T} & 2\I t_{TN} \sin(k_x/2) & 2\I t_{TN}  \sin(k_x/2)\\
	-2\I t_{TN} \sin(k_x/2) & -2t_{T} \cos(k_x) +\epsilon_{T} & 0 \\
	-2\I t_{TN} \sin(k_x/2) & 0  & -2t_{N} \cos(k_x) +\epsilon_{N} \\
	\end{pmatrix} \psi_k,\\
}
\end{widetext}
with the spinor $\psi_k^{\dagger}=(c_{T,1,k}^{\dagger},c_{T,2,k}^{\dagger},c_{N,k}^{\dagger}).$

The main approximation is to neglect the hopping between two adjacent Ta-Ni-Ta chains, which is much smaller $t_{inter}=0.035$ eV than the energy of the gap analyzed in this study. Neglecting the latter, allow us to make a unitary transformation into the bonding-antibonding combination of the Ni-Ta hybridization as $c_{\pm,k}=\sqrt{1/2} [c_{T,1,k} \pm c_{T,2,k}]$. This leads to a decoupling of the antibonding combination of the Ni orbitals from the Ta orbital and rescaling the effective hopping as $t_{TN}\rightarrow \sqrt{2}t_{TN}.$ The same transformation on the level of the interband density-density interactions leads to 
\beq{
	V  n_{N,i} [ n_{T,1,i} + n_{T,2,i}]= V n_{N,i} [n_{-,i}+ + n_{+,i}]  
}
and similar for the electron-phonon interaction. The transformation  separates the three band problem into two blocks of two orbital problems. In the main text and the rest of SM, we have marked Ni band as band 0 and the combination of Ta bands as 1.

We define the order parameter as the  local component of the hybridization between bands $\Delta= \sum_k \langle c_{+,k}^{\dagger} c_{N,k}\rangle$. In Ref.~\onlinecite{mazza2020}, the rather proposed an even combination of hybridization between the combination of Ta and Ni chains as
$\Delta_{+}=\frac{1}{N}\sum_{i}\langle c_{+,i}^{\dagger} c_{N,i} +c_{+,i+1}^{\dagger} c_{N,i} \rangle=2\sum_k \cos(k) \langle c_{+,k}^{\dagger} c_{N,k}\rangle.$ The two definitions of the order parameter  encode essentially the same information as and we have explicitly checked that conclusions in Fig.~\ref{Fig:theo2} only change quantitatively, but not qualitatively.

\subsection{Theoretical photoemission spectrum}
The theoretical photoemission  spectrum~\cite{freericks2009,eckstein2008} is evaluated as
\bsplit{
&I(\omega,k,t_p)=\\&-\text{Im}\left[\sum_{\alpha=0}^1\int dt' S(t') S(t)\exp^{i \omega(t'-t)}G^<_{k,\alpha\alpha}(t',t)\right],}
where $S(t)=\exp(-t^2/\tilde\delta^2)$ is the envelope of the probe pulse with the duration $\tilde\delta=100$ fs centered around delay time $t_p$ and the lesser component of the Green's function is defined as $G^<_{k,\alpha\alpha}(t',t)=\I\langle c^{\dagger}_{k,\alpha}(t') c_{k,\alpha}(t)\rangle$. 

\subsection{Method}
Our objective is to describe the strong lifetime response to photoexcitation while keeping the electron and phonon interactions on the same level of consideration. The nonequilibrium dynamics will be treated within the Keldysh formalism, which is a nonperturbative approach in the strength of the electric field. It is useful to introduce the spinor
\beq{
	\Psi_{k}\equiv 
\begin{pmatrix}
c_{0,k}  \\
c_{1,k}  \\
\end{pmatrix}
}
and the corresponding 2$\times$2  Greens function is given by 
\beq{
	G_k(t,t')=-\I \green{\Psi_k}{\Psi_k},
}
which is determined from a solution of the Dyson's equation. For later usage we will introduce the single-particle density matrix $\rho_{ij,k}=\ave{\Psi^{\dagger}_{j,k} \Psi_{i,k}}$ and its local component
$\rho_{ij,\text{loc}}=\sum_k \rho_{ij,k}.$ The occupation of the valence~(conduction) band is therefore given by $n_0=\rho_{00,\text{loc}}~(n_1=\rho_{11,\text{loc}}).$  The excitonic order parameter $\phi$ is determined by the off-diagonal and local component of the single-particle density matrix $\phi=\rho_{01,\text{loc}}.$

The approximation used is determined by the self-energy and we will treat the Coulomb interaction within the 2nd Born approximation, which includes inelastic scattering and captures the  heating effects.  For the Coulomb interaction,  the Hartree term is given by
\beq{
	\Sigma^H_{ij,k}(t)=\delta_{ij} V \rho_{\bar i\bar i,\text{loc}}(t),
}
where the overline marks the opposite band. The Fock term is given by
\beq{
	\Sigma^F_{ij,k}(t)= V\rho_{ij,\text{loc}}(t).
}
The coupling with phonons will be described on the Hartree-Fock level with the instantaneous term 
\beq{
	\Sigma^{ph}_{ij,k}(t)= g X(t) \delta_{|i-j|=1},
}
and the phonon motion is described by a system of differential equations for the phononic distortion $X$ and the corresponding momentum $P$
\bsplit{
	&\partial_t P(t)=\omega_0 P(t),\\  &\partial_t X(t)=-\omega_0 X(t) -2 g (\rho_{01,\text{loc}}+\rho_{10,\text{loc}}).
}
As the Hartree and Fock self-energies are instantaneous, it is useful to redefine the single particle Hamiltonian as $h_k^{HF}=h_k+\Sigma^H_k+\Sigma_k^F$ and introduce the renormalized free propagator as
\beq{
	G_k^{HF}=(\I\partial_t-h_k-\Sigma^H-\Sigma^F_k-\Sigma^{ph}_{k})^{-1}.
}
In the 2nd Born approximation, the correlation self-energy is given by the expansion in the interaction strength which is then re-summed using the Dyson equation for the diagrammatic representation of the self-energy. The analytical expression for the self-energy is given by
\bsplit{
\Sigma^{2B}_{ij,q}(t,t')=&-\I V^2 \sum_k  G_{\bar j\bar i,k}(t',t) \\
&\left [\chi_{i\bar j,k+q}(t,t')- \chi_{\bar i\bar j,k+q}(t,t')\right]},
\label{Eq:Sigma}
where we have introduced the charge susceptibility
\beq{
\chi_{ij,q}(t,t')=\I \sum_k G_{i\bar j,q-k}(t,t')G_{\bar i j,k}(t,t').
\label{Eq:Pol}
}

The electron-phonon interaction is treated on the level of the Migdal approximation and given by 
\beq{
\Sigma_{ij,q}^{M}(t,t')=\I \lambda D(t,t') \sum_q G_{\bar i \bar j,q}(t,t'),
}
where $D$ is the bosonic propagator. It is obtained by solution of the bosonic Dyson equation
\beq{
D(t,t')=D_0(t,t')+D_0(t,t')\ast \Pi^{\text{Ph}}(t,t')\ast D(t,t'), 
}
where $\ast$ denotes the convolution in time and the matrix multiplication in the orbital space, the $D_0$ is the free phonon propagator and the phonon polarization bubble is $\Pi_{ij,q}^{\text{Ph}}(t,t')=\I\sum_{k}G_{i\bar j,k}(t,t') G_{\bar i j,q-k}(t',t).$

We included a weakly coupled phononic bath with interband coupling in the form
$H_{\text{tot}}=H+\sqrt{\lambda_1}\sum_{i} \tilde X_i \left[c_{0,i}^{\dagger} c_{0,i} + c_{1,i}^{\dagger} c_{1,i}\right] + \sum_i \frac{1}{2}\left[ X_{i}^2 + \frac{1}{\tilde \omega_0^2}\dot X_{i}^2\right]$
to the system so that photo-excited electrons~(holes) can quickly relaxed to the lower~(upper) edge of the conducting~(valence) band. The inclusion of the bath is of a particular importance in the lattice driven situation, where the electronic relaxation is suppresses due to the interband nature of lattice modes and low frequencies of them.  We treat the lattice at the lowest order self-energy expansion~(without selfconsistency)
\beq{\Sigma^{\text{Bath}}_{ii,\text{loc}}(t,t')=\lambda_1 G_{ii,\text{loc}}(t,t') \ast \tilde D_0(t,t'),}
where $\tilde D_0$ is the free propagator for the bath phonon mode $\tilde X_i.$ In all calculations, we choose frequency of the bath mode to be $\tilde \omega_{0}=0.33$ eV and the electron-bath couplings $\lambda_1=0.06$ eV. The inclusion of such a weakly coupled bath leads to qualitatively similar dynamics in all parameter regimes for photo-excited electrons~(holes), which are gathered at the lower~(upper) edge of the conducting~(valence) band.

Finally, the electronic Greens function is obtained from the solution of the electronic Dyson equation
\beq{
	G_k=G_k^{HF} + G_k^{HF}\ast[\Sigma_k^{2B}+\Sigma_k^{M}+\Sigma^{\text{Bath}}]\ast G_k.
}
We solve this problem numerically using the library NESSi employing an efficient MPI parallelization over the momentum points~\cite{schuler2020nessi}.

\subsection{Experimental data analysis} 

\begin{figure}[t]
\includegraphics[width=1.0\linewidth]{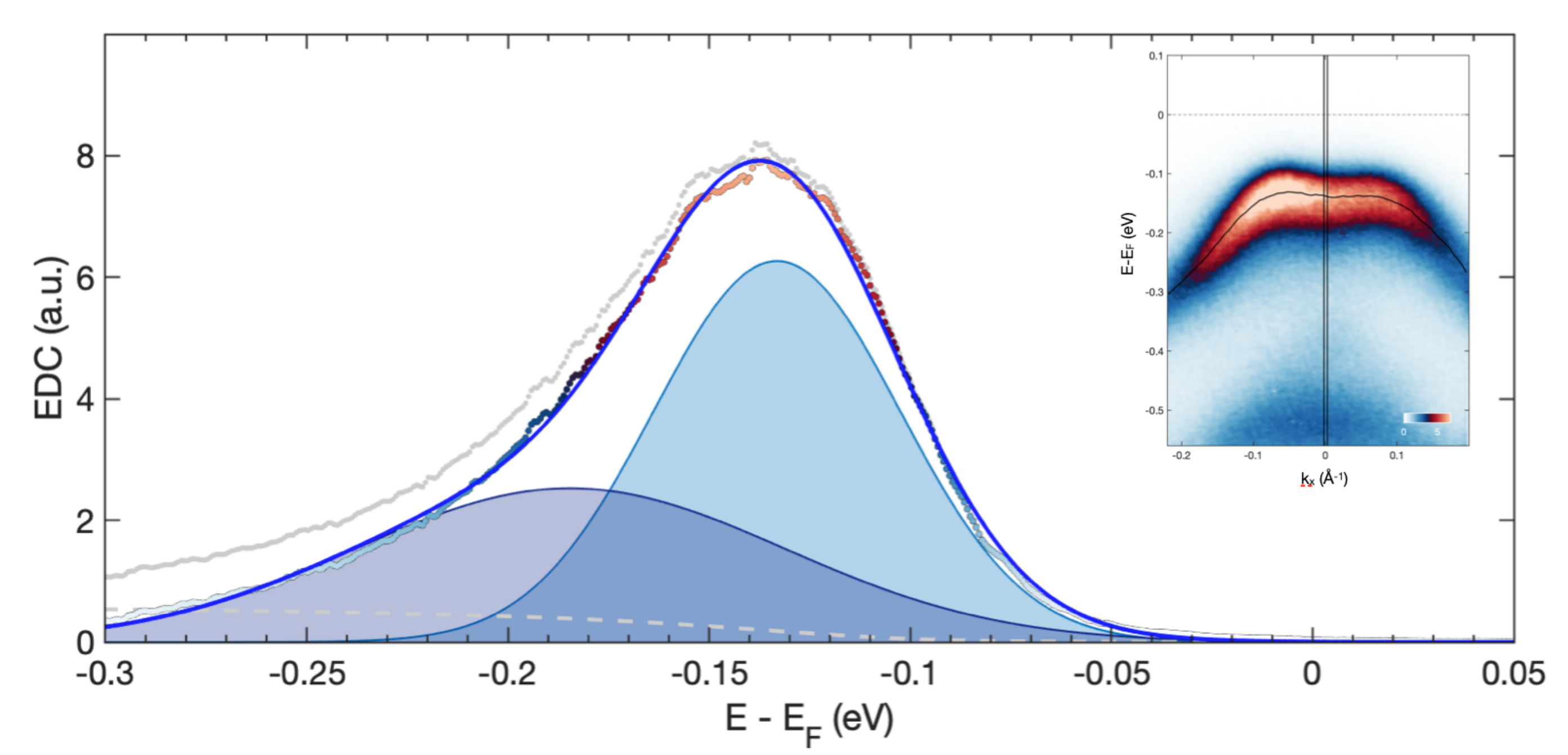}
\caption{Momentum-integrated EDC extracted from $\Gamma$ (highlighted by the black rectangle in the ARPES spectrum inset) before (grey line) and after (coloured circles) the Shirley background (grey dashed line) subtraction. The fit (blue line) is composed of the sum of two Gaussian functions (shaded regions), and is used to extract the FWHM and peak position of the EDC.}
\label{Fig:EDCSM}
\end{figure}

The analysis discussed pertains to the experimental observations of the upper valance band dynamics, and is thus limited to this region. EDCs were obtained by integrating 0.0048 $A^{-1}$ wide momentum intervals across all k-values acquired, and a Shirley background (grey dashed line, Fig.~\ref{Fig:EDCSM}) was subtracted from each EDC.

In order to extract the FWHM and position of the UVB dispersion, we fit the data to a function that best reproduces the asymmetric spectral feature at each k-point so that we may be certain that we can accurately extract the true width and position of the UVB from this fit. The functional form we choose consists of the sum of two Gaussian functions, which fit to the Shirley-background subtracted EDC data. The UVB position (denoted E$_0$ (k$_x$,t) in the main text) is defined by the energy (E - E$_F$) corresponding to the fit maximum; the full width at half maximum (denoted FWHM(k$_x$,t) in the main text) is defined as the difference between the two energy positions that correspond the half-maximum points. The difference quantities presented in the main text ($\Delta$FWHM(k$_x$,t), E$_0$(k$_x$,t)) were determined by subtracting the mean equilibrium FWHM and peak position (calculated using the time-delays between t = -700 fs and t = -500 fs). 

\subsection{Experimental photodoping estimate}
Experimental estimates for the approximate photodoping induced by a 1.55 eV optical pump polarized along the a-axis, at a repetition rate of 250 kHz, were determined using the real optical conductivity and dielectric constant obtained from published results~\cite{larkin2017}. The reflectivity $R$ of the sample with a real $n$ (imaginary $\kappa$) refractive index of 3.09 (2.56) for light incident at 45$^{\circ}$ with respect to the horizontal plane of incidence was calculated using the Fresnel equations. 

The number of photons absorbed at a depth of 1 nm (formula unit) is given by the following expression 
\beq{
	N = N_0 (1-R)[(1-e^{-\alpha d}) - (1 - e^{-\alpha(d - 1nm)})],
}
where $N_0$ is the number of incident photons per pulse (250 kHz repitition rate),  $\alpha = 2\omega k/c$ is the absorption coefficient, and $d$ is the penetration depth, which is estimated to be 30 nm from the mean free path of photoelectrons with a kinetic energy of 2 eV. Assuming a quantum efficiency of one, the number of photoexcited electrons per formula unit for the the indecent pump fluences measured in this experiment are presented in Table ~\ref{Table:t1}. 

\begin{table}
 \begin{tabular}{||c | c||} 
 \hline
 Incident Fluence & Effective Photodoping \\ ($\mu$J/cm$^2$) & (electron/formula unit)  \\ [0.75ex] 
 \hline\hline
 160 & 0.032 \\ 
 \hline
 50 & 0.01  \\
 \hline
 26 & 0.005\\
 \hline
\end{tabular}
\caption{Experimental estimate of the effective photodoping induced by a 1.55 eV pump pulse.}
 \label{Table:t1}
\end{table}

\begin{figure}[t]
\includegraphics[width=1.0\linewidth]{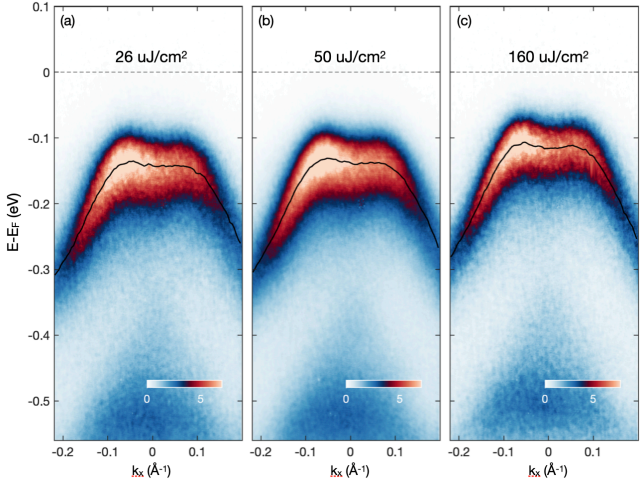}
\caption{TR-ARPES spectrum acquired at a base sample temperature of 90 K prior to pump-probe overlap (t = -600 fs) for incident pump fluences of (a) 26 $\mu$J/cm$^2$, (b) 50 $\mu$J/cm$^2$, and (c) 160 $\mu$J/cm$^2$}
\label{Fig:SMspec2}
\end{figure} 

\subsection{Heating due to finite repetition rate}

The UVB of TNS has a known temperature dependence as reported in~\cite{bandshiftTang}, showing an upward shift in the valance band towards the Fermi level with increasing sample temperature, where it eventually saturates as it approaches the transition temperature. Our TRARPES spectra were acquired with a repetition rate of 250 kHz at a base temperature of 90 K. Due to the high repetition rate of the pump, laser-induced heating was observed even before the arrival of the pump-pulse (t $<$ 0), and became more pronounced for fluences higher than 26 $\mu$J/cm$^2$. This resulted in an observed upward shift in the pre-photoexcitation spectrum to lower binding energies with an increase in incident pump-fluence, as shown in Fig.~\ref{Fig:SMspec2}. The position of the UVB at $\Gamma$ (E - E$_F$ =  -0.142 eV) for the dataset acquired with an indicent pump fluence of 26 $\mu$J/cm$^2$ is consistent with the value reported in~\cite{bandshiftTang} for a sample temperature of 90 K within error. It is difficult to extract the effective electronic temperature prior to the pump arrival due to laser-heating, so we extract the band-gap renormalization for the different incident pump fluence values relative to the equilibrium position of the low-fluence data at a base temperature of 90 K.

As the repetition time is much longer than the nonequilibrium dynamics induced, we can model the gap reduction by an enhanced temperature of the system. Theoretically, we adjust the temperature such that the gap show the same reduction as the experimental one with respect to the low-temperature regime. In Table.~\ref{Table:t2}, we have gathered the experimental bandgap renormalization due to the heating with corresponding theoretical temperatures. In the electron driven case, we could only stabilize solution with lowest temperature at $T=150$ K. The equilibrium transition temperature are overestimated in theoretical calculations are corresponds to $T_C=400$ K for the electronic driven situation~($\lambda/V=0.03$) and $T_C=550$ K for the lattice driven case~($\lambda/V=7.0$).

\begin{table*}
 \begin{tabular}{||c | c | c | c | c||} 
 \hline
 Incident Fluence & Eq. UVB Position & Bandgap Renorm. & Theory T  & Theory T  \\ ($\mu$J/cm$^2$) & E - E$_F$ (eV) & (meV) &  ($\lambda/V=0.03$) (K) & ($\lambda/V=7.0$) (K) \\ [0.75ex]
 \hline\hline
 160 & -0.117 & 25 & 330 & 250  \\ 
 \hline
 50 & -0.139 &  3-5 & 220 & 165 \\
 \hline
 26 & -0.142 & 0 & 150 & 90 \\
 \hline
\end{tabular}
\caption{Bandgap renormalization (relative to UVB position at $\Gamma$ for the incident pump fluence of 26 $\mu$J/cm$^2$) before the pump pulse due to heating and theoretical effective temperatures in the electronic~($\lambda/V=0.03$) and the lattice driven~($\lambda/V=7.0$) case.}
\label{Table:t2}
\end{table*}
\end{document}